\documentclass{article}
\usepackage{vmargin}
\usepackage{xcolor}
\usepackage{amsmath,amsfonts}
\usepackage{graphicx}
\usepackage{accents}
\makeindex

\begin{document}

\begin{center}
\textbf{\large{Electromagnetic waves in photonic crystals: laws of dispersion,\\[2mm]
causality and analytical properties}}\\[8mm]

{Boris Gralak$^\dagger$, Maxence Cassier, Guillaume Dem{\'e}sy and S{\'e}bastien Guenneau\\[3mm]
\textit{Aix Marseille Univ, CNRS, Centrale Marseille\\ 
Institut Fresnel, Marseille, France} \\[3mm]
$^\dagger$ 
{\color{blue}{\texttt{\textup{boris.gralak@fresnel.fr}}}}
}\\[10mm]
\end{center}

\begin{abstract}
Photonic crystals are periodic structures which prevent light propagation along
one or more directions in certain frequency intervals. Their band spectrum is 
usually analyzed using Floquet-Bloch decomposition. This spectrum is located on 
the real axis, and it enters the complex plane when 
absorption and dispersion is considered in the dielectric permittivity of 
material constituents. Here, we review fundamental definition and properties 
of dispersion law and group velocity in photonic crystals and we illustrate them with 
numerical examples. \end{abstract}


%
%
\def\comment{\color{blue}}
\newcommand{\dbtilde}[1]{\accentset{\scalebox{0.8}{$\approx$}}{#1}}
\def\x{\text{\bfseries\sffamily\textit{x}}}
\def\xp{\text{\bfseries\sffamily\textit{x}}_{\!\text{\sffamily\textit{p}}}}
\def\xs{\text{\sffamily\textit{x}}}
\def\y{\text{\bfseries\sffamily\textit{y}}}
\def\ys{\text{\sffamily\textit{y}}}
\def\k{\text{\bfseries\sffamily\textit{k}}}
\def\kp{\text{\bfseries\sffamily\textit{k}}_{\!\text{\sffamily\textit{p}}}}
\def\u{\text{\bfseries\sffamily\textit{u}}}
\def\e{\text{\bfseries\sffamily\textit{e}}}
\def\a{\text{\bfseries\sffamily\textit{a}}}
\def\v{\text{\bfseries\sffamily\textit{v}}}
\def\lat{L}
\def\cell{V}
\def\p{\text{\bfseries\sffamily\textit{p}}}
\def\K{\text{\bfseries\sffamily\textit{K}}}
\def\K{\text{\bfseries\textit{K}}}
\def\K{\text{\bfseries\sffamily\textit{a}}^\ast}
\def\Bri{B}
\def\reclat{L^\ast}
\def\En{\mathcal{E}}

\def\E{\text{\bfseries\sffamily\textit{E}}}
\def\P{\text{\bfseries\sffamily\textit{P}}}
\def\B{\text{\bfseries\sffamily\textit{B}}}
\def\D{\text{\bfseries\sffamily\textit{D}}}
\def\A{\text{\bfseries\sffamily\textit{A}}}
\def\J{\text{\bfseries\sffamily\textit{J}}}
\def\F{\text{\sffamily\textit{F}}}
\def\t{\text{\sffamily\textit{t}}}
\def\t{t}
\def\s{\text{\sffamily\textit{s}}}
\def\nui{\text{\sffamily\textit{$\nu$}}}
\def\f{\text{\sffamily{f}}}
\def\c{\text{\sffamily\textit{c}}}
\def\c{c}
\def\ks{\text{\sffamily\textit{k}}}
\def\rot{\boldsymbol{\nabla \times }}
\def\dt{\partial_{\t}}
\def\hatLE{\hspace*{0.35mm}\widehat{\hspace*{-0.35mm}\text{\bfseries\sffamily\textit{E}}\hspace*{0.35mm}}\hspace*{-0.35mm}}
\def\LJ{\,\widehat{\!\text{\bfseries\sffamily\textit{J}}\,}\!}
\def\LD{\,\widehat{\!\text{\bfseries\sffamily\textit{D}}\,}\!}
\def\E{\text{\bfseries\textit{E}}}
\def\hE{\,\widehat{\!\boldsymbol{E}}}
\def\P{\text{\bfseries\textit{P}}}
\def\B{\text{\bfseries\textit{B}}}
\def\K{\text{\bfseries\textit{K}}}
\def\H{\text{\bfseries\textit{H}}}
\def\hH{\,\widehat{\!\boldsymbol{H}}}
\def\D{\text{\bfseries\textit{D}}}
\def\A{\text{\bfseries\textit{A}}}
\def\J{\text{\bfseries\textit{J}}}
\def\F{\text{\textit{F}}}
\def\F{\text{\bfseries\textit{F}}}
\def\X{\text{\bfseries\textit{X}}}
\def\V{\text{\bfseries\textit{V}}}
\def\hF{\,\widehat{\!\boldsymbol{F}}}
\def\LE{\text{\bfseries\sffamily\textit{{E}}}}
\def\LEs{\text{\sffamily\textit{{E}}}}
\def\LP{\text{\bfseries\sffamily\textit{P}}}
\def\LB{\text{\bfseries\sffamily\textit{B}}}
\def\LH{\text{\bfseries\sffamily\textit{H}}}
\def\LD{\text{\bfseries\sffamily\textit{D}}}
\def\LA{\text{\bfseries\sffamily\textit{A}}}
\def\LJ{\text{\bfseries\sffamily\textit{J}}}
\def\LF{\text{\sffamily\textit{F}}}
\def\LS{\text{\sffamily\textit{S}}}
\def\vS{\text{\bfseries\sffamily\textit{S}}}
\def\ep{\varepsilon}
\def\om{\omega}
\def\x{\text{\bfseries\sffamily\textit{x}}}
\def\y{\text{\bfseries\sffamily\textit{y}}}
\def\k{\text{\bfseries\sffamily\textit{k}}}

\def\rot{\partial_{\x} \times\!}
\def\dint{\displaystyle\int}
\def\R{\mathbb{R}}
\def\C{\mathbb{C}}
\def\Z{\mathbb{Z}}
\def\Proj{\mathsf{P}}
\def\He{\mathsf{H}}
\def\Res{\mathsf{R}}
\def\Re{\mathsf{R}_e}
\def\G{\mathsf{G}}


\section{Introduction\label{sec1}}

Photonic crystals are periodic electromagnetic structures that have been originally introduced by 
Eli Yablonovith \cite{Yab87} and Sajeev John \cite{Sajeev87} in order to inhibit the 
spontaneous emission \cite{Bykov,Yab87} and obtain strong localization of photons\cite{Sajeev87}. 
The original idea, based on an analogy with solid states Physics \cite{Kittel}, 
was to use the periodic modulation in two or three dimensions of a lossless dielectric permittivity 
to open photonic bandgaps, i.e. ranges of frequencies where for which the electromagnetic radiation cannot 
propagate \cite{Yab91}. If an excited atom 
is embedded in such a periodic medium and if its energy level corresponds to a frequency of the bandgap, 
then photons cannot be radiated. Therefore photons can be strongly localized \cite{Sajeev87} and
the spontaneous emission can be inhibited \cite{Bykov}. 

Hence, an important challenge of photonic crystals topic was to obtain in three-dimensions 
at optical wavelengths a full 
photonic bandgap (i.e. light is disallowed to propagate along all directions) sufficiently 
robust to the fabrication imperfections. The most 
promising structures have probably been the photonic crystals produced using colloidal 
suspensions \cite{RSS95,BRW97}, layer-by-layer semiconductor industry technique 
\cite{LFH+98,FL99,NTYC00,LFL+01} and inverse opal synthesis \cite{BCG+00,VBS+01}. 
Nevertheless, the fabrication of such three-dimensional 
structures remains difficult to proceed, notably in comparison with the fabrication of 
two-dimensional photonic crystals for which the semiconductor techniques can be directly 
transposed to etch membranes or slabs on substrate \cite{Krauss}. 

The ability of two-dimensional photonic crystals to forbid the propagation of the 
electromagnetic field has been exploited to guide light in microstructured 
optical fibres \cite{knight96,white,Kuhlmey,guenneau03a,nicolet04a,zolla05a} and 
planar structures in integrated optics \cite{Weisbuch}. 
In photonic crystal fibers the photonic bandgap allows the guiding of light in air 
or vacuum, thus enabling to enhance the power of the guided light. In integrated optics, 
the objective was to obtain optical circuits with both reduced dimensions and
a reduction of the radiation losses \cite{Jamois}. Furthermore, the two-dimensional 
photonic bandgap has been used to design cavities with high quality factor 
\cite{Noda00,Noda03}, with applications to the enhancement of the efficiency of 
light sources and sensors. In that case, the enhancement of the 
emission of photons and the electromagnetic local density of states is based on the 
existence of bandgaps, i.e. the absence of photonic modes for certain frequency 
ranges in photonic crystals. 

The photonic bands themselves can be also exploited to obtain a fine control 
of the emission and propagation of electromagnetic waves. In addition to its enhancement, 
the emission of electromagnetic waves can be channelled around specific directions as soon 
as the photonic bands are restricted to the corresponding ranges of 
wavevectors \cite{EGT02}, with applications to directive antennas \cite{Enoch02}. 
The propagation of electromagnetic waves is governed by the photonic bands providing 
the dispersion law and the group velocity \cite{Yeh79,chap10}. The richness of the 
dispersion law can lead to an enhanced dispersive effect or, conversely, to a self 
guiding effect \cite{EGT03}, and to exotic refraction properties like ultra-refraction 
and negative refraction \cite{Gralak00}. In particular, negative refraction from 
photonic crystals \cite{PRL06,ARG09,APL10} can be considered as an alternative to 
negative index from metamaterials \cite{Pendry00,Smith00,Smith04} since it is 
not spoiled by absorption. 

All the aforementioned effects and applications are governed by the photonic bands 
and gaps which are totally determined by the relationship between the frequency $\omega$ and 
the wavevector $\boldsymbol{k}$, namely the dispersion law.
This chapter will be devoted to this relationship  including the last 
developments with dispersion and absorption. After the presentation of Maxwell's equations 
in photonic crystals in section \ref{sec2}, the Floquet-Bloch decomposition is introduced 
in section \ref{sec3}. 
It is shown that this Floquet-Bloch decomposition is a unitary transform which is 
specially adapted to partial differential equations with periodic 
coefficients since it commutes with multiplicative operator by periodic functions. 
Then, the dispersion law $\om(\k)$ is introduced in section \ref{sec4} and it is shown 
that the group velocity $\partial_\k \om (\k)$ governs the propagation of the 
electromagnetic field. In section \ref{sec5}, numerical calculations of the dispersion 
law are presented in the case of two-dimensional photonic crystals. In addition, 
the effect of the effective anisotropy on the propagation of the electromagnetic 
field is numerically illustrated in two-dimensional photonic crystals. In section \ref{sec6} 
the dispersion law is extended to dispersive and absorptive photonic crystals and 
numerical calculation of the complex spectrum of Bloch resonances are provided 
for two-dimensional photonic crystals made of a Drude metal. Finally, the analytic 
nature of the dispersion law is discussed in section \ref{sec7}. 

\section{Maxwell's equations in photonic crystals\label{sec2}}

In this chapter, different bases are used: $(\e_1,\e_2,\e_3)$ is 
an orthonormal basis; $(\a_1,\a_2,\a_3)$ is the basis defining the lattice 
associated with the photonic crystal, hence it need not be orthonormal; 
and $(\K_1,\K_2,\K_3)$ is the basis defining the reciprocal lattice. 
Every vector $\x$ in $\R^3$ (respectively in $\C^3$) of the physical space 
is described by three components $\xs_1$, 
$\xs_2$ and $\xs_3$ in $\R$ (respectively in $\C$).
 
We start with macroscopic Maxwell's equations in linear, dispersion-free dielectric media:
\begin{equation}
\begin{array}{l}
\rot \E(\x,t) = - \mu_0 \partial_t \H(\x,t) \, , \\[2mm] 
\rot \H(\x,t) = \ep(\x) \partial_t \E(\x,t) + \J(\x,t) \, , 
\end{array}
\label{Me}
\end{equation}
where $\E(\x,t)$ and $\H(\x,t)$ are the electric and magnetic 
fields, $\J(\x,t)$ is the current source density, 
$\rot$ is the curl operator, $\mu_0$ is the vacuum permeability and $\ep(\x)$ is 
the dielectric permittivity. The dielectric permittivity $\ep(\x)$ in photonic crystals 
is generally considered as a frequency-independent function taking real and positive values 
greater than the one of the vacuum permittivity $\ep_0$. Indeed, such functions can describe  
lossless dielectric materials which are good candidates to obtain bandgap in photonic 
crystals \cite{JMW95}, while the presence of absorption implies the absence of bandgaps 
\cite{Tip00}. In this chapter, the case of dispersive and absorptive permittivity 
is addressed in the section \ref{sec6}. In the other sections, one assumes that
the photonic crystal is neither dispersive nor dissipative.

Let $\a_1,$ $\a_2,$ and $\a_3$ be the linearly-independent and non-vanishing 
vectors of $\R^3$ defining the unit cell $\cell$ of the periodic photonic crystal:
\begin{equation}
\cell = \big\{ \, \x = x_1 \a_1 + x_2 \a_2 + x_3 \a_3 \, \big| \, x_1,x_2,x_3 \in [0,1] \, \big\} \, .
\label{cell}
\end{equation}
Then, the lattice $\lat$ associated with the photonic crystal is 
\begin{equation}
\lat = \big\{ \, \a = p_1 \a_1 + p_2 \a_2 + p_3 \a_3 \, \big| \, p_1,p_2,p_3 \in \Z \, \big\} \, 
\label{lat}
\end{equation}
and the permittivity $\ep(\x)$ determining the geometry of the crystal is invariant 
under the set of translations by the vectors of the lattice:
\begin{equation}
\ep(\x + \a) = \ep(\x) \, , \qquad  \x \in \R^3 \, , \, \a \in \lat \, .
\label{per}
\end{equation}
The basis $(\K_1,\K_2,\K_3)$ of the reciprocal lattice is defined such that 
$\K_i \cdot \a_j = 2 \pi \delta_{ij}$ with $\delta_{ij}$ the Kronecker symbol 
($\delta_{ij} = 1$ if $i=j$ and $\delta_{ij} = 0$ otherwise):
\begin{equation}
\K_1 = \dfrac{2 \pi}{A} \, \a_2 \times \a_3 \, , \qquad 
\K_2 = \dfrac{2 \pi}{A} \, \a_3 \times \a_1 \, , \qquad 
\K_3 = \dfrac{2 \pi}{A} \, \a_1 \times \a_2 \, , 
\label{Kij}
\end{equation}
where $A = | \, (\a_1 \times \a_2) \cdot \a_3 \, | \neq 0$ is the volume of the unit cell 
$\cell$. Finally, the reciprocal lattice is defined by
\begin{equation}
\reclat = \big\{ \, \K = p_1 \K_1 + p_2 \K_2 + p_3 \K_3 \, \big| \, p_1,p_2,p_3 \in \Z \, \big\} \, 
\label{reclat}
\end{equation}
and the unit cell $B$ of this reciprocal lattice, or the 
\textbf{first Brillouin zone}, is given by 
\begin{equation}
\Bri = \big\{ \, \k = k_1 \K_1 + k_2 \K_2 + k_3 \K_3 \, \big| \, k_1,k_2,k_3 
\in [-1/2, 1/2] \, \big\} \, .
\label{Bri}
\end{equation}

The following purely electromagnetic quantity, corresponding to the electromagnetic 
energy if the field is in vacuum, is assumed to be finite
for all time $t$:
\begin{equation}
\En(t) = \dfrac{1}{2} \displaystyle\int_{\R^3} \, d\x \, \big[ 
\ep_ 0 \E(\x,t)^2 + \mu_0 \H(\x,t)^2 \big] \quad < \infty \, . 
\label{energy}
\end{equation}
This assumption implies that the electromagnetic fields $\E(\x,t)$ and $\H(\x,t)$
are square integrable functions of the position $\x$ with 
well-defined Fourier transforms $\hE(\k,t)$ and $\hH(\k,t)$.

\section{The Floquet-Bloch decomposition\label{sec3}}

The Floquet-Bloch decomposition is a unitary transform adapted to partial 
derivative equations with periodic coefficients \cite{bensoussan78a,Kuc93}. 
This decomposition exploits the invariance of the equations under the group 
of symmetries formed by the set of translations by the vectors $\a$ in the 
lattice $\lat$ (\ref{lat}). In this chapter, it is shown that the Maxwell's 
equations (\ref{Me}) with periodic permittivity $\ep(\x)$ are 
equivalent to a family of similar equations indexed by the Bloch wavevector $\k$ 
and restricted to the unit cell $\cell$. Also, the electromagnetic fields 
$\E(\x,t)$ and $\H(\x,t)$ can be uniquely defined as the superposition of 
Bloch waves indexed by the Bloch wavevector $\k$ spanning the Brillouin 
zone $\Bri$. The arguments supporting these results, based on the Fourier 
transform and Fourier series are briefly presented and then are concluded 
by a summary on the Floquet-Bloch decomposition. \\

\noindent
\textbf{From the Fourier analysis to the Floquet-Bloch 
decomposition.}
The Fourier transforms $\hE(\k,t)$ and $\hH(\k,t)$ can be related to the 
fields $\E(\x,t)$ and $\H(\x,t)$ with the following integral expressions: 
for $\F = \E, \H$
\begin{equation}
\hF(\k,t) = \displaystyle\int_{\R^3} d\x \, 
\exp[- i \k \cdot \x] \, \F(\x,t) \, ,
\label{FouF}
\end{equation}
and, conversely, 
\begin{equation}
\F(\x,t) = \dfrac{1}{(2\pi)^3} \displaystyle\int_{\R^3} d\k \, \exp[ i \k \cdot \x] \, \hF(\k,t) \, .
\label{FFou}
\end{equation}
This last expression can be decomposed using the reciprocal lattice:
\begin{equation}
\F(\x,t) = \dfrac{1}{(2\pi)^3} \displaystyle\int_{\Bri} d\k \, \displaystyle\sum_{\K \in \reclat} 
\exp[ i (\k + \K) \cdot \x] \, \hF(\k+\K,t) \, . 
\label{Fdecomp}
\end{equation}
Let $\F_{\!\#}(\x,\k,t)$ denote the series under the integral:
\begin{equation}
\F_{\!\#}(\x,\k,t)= \dfrac{1}{(2\pi)^3} \displaystyle\sum_{\K \in \reclat} 
\exp[ i (\k + \K) \cdot \x] \, \hF(\k+\K,t) \, .
\label{Fper}
\end{equation}
This function appears to be periodic of $\k$, i.e. invariant under translations of 
vectors $\K$ in the reciprocal lattice $\reclat$. Hence, 
it can be expanded as the Fourier series
\begin{equation}
\F_{\!\#}(\x,\k,t) = \displaystyle\sum_{\a \in \lat} 
\exp[- i \k \cdot \a] \, \dfrac{A}{(2 \pi)^3} \displaystyle\int_{\Bri} d\k' \, 
\exp[ i \k' \cdot \a] \, \F_{\!\#}(\x,\k',t) \, .
\label{Fseries}
\end{equation}
where it has been used that $(2 \pi)^3 /A = | (\K_1 \times \K_2) \cdot \K_3 |$ is the 
volume of the first Brillouin zone $\Bri$. Replacing $\F_{\!\#}(\x,\k',t)$ by its series 
expression (\ref{Fper}), 
the coefficients of the Fourier series become (up to the factor $A/(2\pi)^3$)
\begin{equation}
\begin{array}{l}
\displaystyle\int_{\Bri} d\k' \, \exp[ i \k' \cdot \a] \, \F_{\!\#}(\x,\k',t) \\[4mm]
\qquad \qquad \qquad = 
\displaystyle\int_{\Bri} d\k' \, \exp[ i \k' \cdot \a] \, 
\dfrac{1}{(2\pi)^3} \displaystyle\sum_{\K \in \reclat} 
\exp[ i (\k' + \K) \cdot \x] \, \hF(\k'+\K,t) \, \\[4mm]
\qquad \qquad \qquad = 
\displaystyle\int_{\Bri} d\k' \, 
\dfrac{1}{(2\pi)^3}\displaystyle\sum_{\K \in \reclat} 
\exp[ i (\k' + \K) \cdot (\x + \a)] \, \hF(\k'+\K,t) \, , 
\label{coeff}
\end{array}
\end{equation}
where we used that $\exp[i \, \K \!\cdot \a] = 1$. From 
(\ref{Fdecomp}), the coefficients in (\ref{Fseries})--(\ref{coeff}) are 
\begin{equation}
\displaystyle\int_{\Bri} d\k' \, \exp[ i \k' \cdot \a] \, \F_{\!\#}(\x,\k',t) 
= \F(\x + \a,t) \, .
\label{coeff2}
\end{equation}
Hence, combining (\ref{Fdecomp}), (\ref{Fper}), (\ref{Fseries}) and (\ref{coeff2}), 
we deduce that the function $F(\x,t)$ can be written as the 
superposition over the the first Brillouin zone 
\begin{equation}
\F(\x,t) = \displaystyle\int_{\Bri} d\k \, \F_{\!\#}(\x,\k,t) \, , 
\label{decomp}
\end{equation}
with components
\begin{equation}
\F_{\!\#}(\x,\k,t) = \dfrac{A}{(2 \pi)^3} 
\displaystyle\sum_{\a \in \lat} \exp[- i \k \cdot \a] \, \F(\x + \a,t) \, .
\label{FperB}
\end{equation}
The linear transformation (\ref{FperB}) that defines $\F_{\!\#}(\x,\k,t)$ in term 
of $\F(\x,t)$ is usually referred in the literature\cite{Kuc93} as the Floquet-Bloch transform 
of the function $\F(\x,t)$, whereas the equation (\ref{decomp}) gives the expression of the 
inverse of this transform. In addition, a Parseval identity between the norms of $\F(\x,t)$ and 
$\F_{\!\#}(\x,\k,t)$, in the sense of square 
integrable functions, can be derived. 
Starting with the square of the norm of $\F(\x,t)$, the integral over the 
variable $\x$ is decomposed according to the $\lat$-lattice:
\begin{equation}
\displaystyle\int_{\R^3} d\x \, 
\big| \F(\x,t) \big|^2 = \displaystyle\int_{\cell} d\x \, 
\displaystyle\sum_{\a \in \lat} 
\big| \F(\x + \a,t) \big|^2 \, .
\label{iso1}
\end{equation}
From (\ref{FperB}), the function $\F_{\!\#}(\x,\k,t)$ is periodic of 
$\k$ and the coefficients of its series expansion are $\F(\x + \a,t)$. 
Hence, the square of the norm of $\F_{\!\#}(\x,\k,t)$ is 
\begin{equation}
\displaystyle\frac{A}{(2 \pi)^3}
\displaystyle\int_{\Bri} d\k \, \big| \F_{\!\#}(\x,\k,t) \big|^2 
= \displaystyle\sum_{\a \in \lat} 
\big| \F(\x + \a,t) \big|^2 \, ,
\end{equation}
and the identity (\ref{iso1}) becomes
\begin{equation}
\displaystyle\int_{\R^3} d\x \, 
\big| \F(\x,t) \big|^2 = \displaystyle\frac{A}{(2 \pi)^3}
\displaystyle\int_{\cell} d\x \, 
\displaystyle\int_{\Bri} d\k \, \big| \F_{\!\#}(\x,\k,t) \big|^2  \, .
\label{iso}
\end{equation}
\textit{Notice that the integrals and the series have been manipulated
formally without particular cautions. Rigorously, it is necessary to first consider 
fields decreasing ``rapidly'' (e.g. in the Schwartz space\cite{RSII}) and then to 
extend the results to all finite energy (square integrable) fields using a density 
argument. }\\

\noindent
\textbf{Summary.}
{The electromagnetic fields $\F(\x,t)$ with finite energy can be 
decomposed as the superposition (\ref{decomp}), over the first Brillouin zone 
$\Bri$, of the Bloch waves $\F_{\!\#}(\x,\k,t)$ given by the equation (\ref{FperB}). 
The Bloch waves $\F_{\!\#}(\x,\k,t)$, defined for all Bloch wavevector $\k$, 
represent the Floquet-Bloch transform of
$\F(\x,t)$. From (\ref{iso}), the Floquet-Bloch transform is an isometry 
and the decomposition (\ref{decomp}) is unique. The Floquet-Bloch transform 
$\F_{\!\#}(\x,\k,t)$ is a $\reclat$-periodic function with respect to the Bloch 
wavevector $\k$ while, with respect to the space variable $\x$, it 
satisfies the Bloch requirement}
\begin{equation}
\F_{\!\#}(\x + \a,\k,t) = \exp[i \k \cdot \a] \F_{\!\#}(\x,\k,t) \, , \qquad 
\a \in \lat \, . 
\label{Bloch}
\end{equation}
It is important to notice that the Floquet-Bloch transform of 
$\ep(\x) \E(\x,t)$ is  (\ref{FperB})
\begin{equation}
\begin{array}{ll}
\big[ \ep \E \big]_{\!\#}(\x,\k,t) & = \displaystyle\frac{A}{(2 \pi)^3}
\displaystyle\sum_{\a \in \lat} 
\exp[- i \k \cdot \a] \, \ep(\x+\a) \, \E(\x + \a,t) \\[4mm]
& = \displaystyle\frac{A}{(2 \pi)^3}
\displaystyle\sum_{\a \in \lat} 
\exp[- i \k \cdot \a] \, \ep(\x) \, \E(\x + \a,t) \\[4mm]
& = \ep(\x) \, \E_{\!\#}(\x,\k,t) \, . \\[4mm]
\end{array}
\end{equation}
In other words, the Floquet-Bloch transform commutes with the multiplicative 
operator by the periodic function $\ep(\x)$. Thus the Floquet-Bloch transform 
appears to be particularly adapted 
to partial differential equations with periodic coefficients. 
The application of the Floquet-Bloch transform to the Maxwell's 
equations (\ref{Me}) leads to the following family of independent equations 
indexed by the Bloch wavevector $\k$ spanning the 
first Brillouin zone $\Bri$:
\begin{equation}
\begin{array}{l}
\rot \E_{\!\#}(\x,\k,t) = - \mu_0 \partial_t \H_{\!\#}(\x,\k,t) \, , \\[2mm] 
\rot \H_{\!\#}(\x,\k,t) = \ep(\x) \partial_t \E_{\!\#}(\x,\k,t) + 
\J_{\!\#}(\x,\k,t) \, .
\end{array}
\label{FBMe}
\end{equation}
Each equation indexed by $\k$ can be solved separately for fields 
$\E_{\!\#}(\x,\k,t)$ and $\H_{\!\#}(\x,\k,t)$ that are square integrable with 
respect to $\x$ on the unit cell $\cell$. Finally, the  
solutions $\E(\x,t)$ and $\H(\x,t)$ of the initial Maxwell's equations 
are retrieved performing the superposition over the first Brillouin 
zone:
\begin{equation}
\E(\x,t) = \displaystyle\int_{\Bri} d\k \, \E_{\!\#}(\x,\k,t) \, , \qquad \qquad 
\H(\x,t) = \displaystyle\int_{\Bri} d\k \, \H_{\!\#}(\x,\k,t) \, .
\label{decompEH}
\end{equation}
The Floquet-Bloch transform appears as the tool to decompose 
periodic equations into a set of equations restricted to the unit cell $\cell$. 
This transform leads also to the introduction of the Bloch wavevector $\k$,
which is the fundamental physical conserved quantity associated with the group 
of symmetries formed by the set of translations of vector $\a$ in $\lat$.

\section{The dispersion law\label{sec4}}

The Maxwell's equations (\ref{Me}) are invariant under any translation with 
respect to the time $t$. That suggests to decompose the equations with respect 
to the time and to consider them in the time-harmonic regime. 

We start with Maxwell's equations (\ref{FBMe}) after the Floquet-Bloch 
decomposition and with the current source $\J_{\!\#}(\x,\k,t)$ set to zero. 
Assuming that a Fourier decomposition with respect to the time can be applied to 
these equations\footnote{\textit{It is stressed that the Fourier 
decomposition with respect to 
the time of equations (\ref{FBMe}) [or (\ref{Me})] is not straightforward when 
the electromagnetic energy is conserved, and it has to be considered 
in the sense of distributions. An alternative way is to perform a Laplace 
transform\cite{GT10} to the equations for a frequency $\om$ with a positive 
imaginary part \cite{JMP17}. Then, the limit Im$(\om) \downarrow 0$ can be 
considered to define the time-harmonic Maxwell's equations (or the Helmholtz 
operator). }}, the following set of equations is obtained:
\begin{equation}
\begin{array}{l}
\rot \LE_{\#}(\x,\k,\om) = i \om \mu_0 \LH_{\#}(\x,\k,\om) \, , \\[2mm] 
\rot \LH_{\#}(\x,\k,\om) = - i \om \ep(\x) \LE_{\#}(\x,\k,\om) 
\, , 
\end{array}
\label{thMe}
\end{equation}
where $\LE_{\#}(\x,\k,\om)$ and $\LH_{\#}(\x,\k,\om)$ are the time-harmonic 
electric and magnetic fields oscillating at the frequency $\om$, with the 
Bloch boundary condition (\ref{Bloch}). 
This set of equations (\ref{thMe}) can be expressed as the eigenvalue problem 
\begin{equation}
M(\x,\k) \, F_{\!\#}(\x,\k,\om) = \om \, F_{\!\#}(\x,\k,\om) \, , 
\label{Shr}
\end{equation}
where
\begin{equation}
F_{\!\#}(\x,\k,\om) = \left[ \begin{array}{l}
\LE_{\#}(\x,\k,\om) \\  \LH_{\#}(\x,\k,\om) \end{array} \right] \, 
\exp[-i \k \cdot \x] \, , 
\label{FperSh}
\end{equation}
is a square integrable periodic function of $\x$ on the unit cell $\cell$ and 
\begin{equation}
\quad M(\x,\k) = \left[ \begin{array}{lr}
0 & i \ep^{-1}(\x) (\partial_{\x} + i \k)\times\! \\  
-i \mu_0^{-1} (\partial_{\x} + i \k)\times\! & 0 \end{array} \right] \, , 
\label{mat}
\end{equation}
is an operator depending on the Bloch wavevector $\k$. The solutions of time-harmonic 
Maxwell equations (\ref{thMe}) without sources are the \textbf{Bloch modes} of the 
photonic crystal. These modes are proportional to the eigenvectors $F_{\!\#}(\x,\k,\om)$ 
of the operator $M(\x,\k)$ acting on the square integrable periodic functions of $\x$ 
on the unit cell $\cell$. The oscillating frequencies $\om$ of the Bloch modes 
are the eigenvalues of the operator $M(\x,\k)$, hence they depend on the 
wavevector $\k$. This relationship $\om(\k)$ defines the \textbf{dispersion law}. 

The dispersion law $\om(\k)$ and the Bloch modes play a fundamental role. 
Indeed, the solutions of Maxwell's equations in photonic crystals can be 
expressed as a superposition (\ref{decompEH}) of Bloch modes. As to the 
dispersion law $\om(\k)$, it provides the relationship between the two physical 
invariant quantities $\om$ and $\k$ resulting from the temporal and spatial 
symmetries. It governs the propagation of the electromagnetic field through the 
\textbf{group velocity} \cite{Yeh79}
\begin{equation}
\v_g = [ \partial_\k \om ] (\k) \, .
\end{equation}
This property can be justified using the following arguments. Let $\X(t)$ be the center 
of the electric field intensity:
\begin{equation}
\X(t) = \dfrac{\displaystyle\int_{\R^3} \, d\x \: \x \E(\x,t)^2 }
{\displaystyle\int_{\R^3} \, d\x \: \E(\x,t)^2} \,  ,
\end{equation}
where $\overline{\E(\x,t)} = \E(\x,t)$ since the time dependent field is real. Using the unitary 
property of the Floquet-Bloch transform, this vector becomes
\begin{equation}
\X(t) = \dfrac{\displaystyle\int_{\cell} \, d\x 
\displaystyle\int_{\Bri} \, d\k \: [\x\E]_{\#}(\x,\k,t) \cdot 
\overline{\E_{\#}(\x,\k,t)} }{\displaystyle\int_{\cell} \, d\x 
\displaystyle\int_{\Bri} \, d\k \: 
\big| \E_{\#}(\x,\k,t) \big|^2}
\end{equation}
Next, the Floquet-Bloch transform of $\x \E(\x,t)$ is derived from the 
expression (\ref{FperB}):
\begin{equation}
[\x\E]_{\#}(\x,\k,t) = \displaystyle\sum_{\a \in \lat} 
\exp[- i \k \cdot \a] \, (\x+\a) \E(\x + \a,t) = (\x + i \partial_{\k} ) \,
\E_{\!\#}(\x,\k,t) \, . 
\end{equation}
Then, it is assumed that each Floquet-Bloch component $\E_{\!\#}(\x,\k,t)$ is 
made of a single time-harmonic Bloch mode{\footnote{If the components $\E_{\!\#}(\x,\k,t)$ contain several Bloch modes,
then a finite sum over the corresponding bands is obtained: note that each band has a different group velocity.}} 
oscillating at the frequency $\om(\k)$: 
$\E_{\!\#}(\x,\k,t) = \LE_{\#}(\x,\k) \exp[- i\om(\k) t]$. Hence the expression 
above becomes
\begin{equation}
[\x\E]_{\#}(\x,\k,t) = \big\{ [ \x + i \partial_{\k} ]  \LE_{\#}(\x,\k) + 
[\partial_{\k} \om](\k) \, t \, \LE_{\#}(\x,\k)  \big\} \exp[- i\om(\k) t] \, ,
\end{equation}
and the vector $\X(t)$ can be written
\begin{equation}
\X(t) = \X_0 + \V \, t \, ,
\label{perturb}
\end{equation}
where the vectors $\X_0$ and $\V$ are time-independent:
\begin{equation}
\X_0 = \dfrac{\displaystyle\int_{\cell} \, d\x 
\displaystyle\int_{\Bri} \, d\k \:  [ \x + i \partial_{\k} ]  \LE_{\#}(\x,\k) 
\cdot \overline{\LE_{\#}(\x,\k)} }
{\displaystyle\int_{\cell} \, d\x \displaystyle\int_{\Bri} \, d\k \: 
\big| \LE_{\#}(\x,\k) \big|^2} \, , \ 
\end{equation}
and 
\begin{equation}
\V = \dfrac{\displaystyle\int_{\cell} \, d\x 
\displaystyle\int_{\Bri} \, d\k \:  [\partial_{\k} \om](\k) \, 
\big| \LE_{\#}(\x,\k) \big|^2 }{\displaystyle\int_{\cell} \, d\x 
\displaystyle\int_{\Bri} \, d\k \: \big| \LE_{\#}(\x,\k) \big|^2} \, .
\end{equation}
Thus the vector $\V$ appears as the velocity of the field intensity center. 
Its expression as the average of the group velocity shows that the latter 
governs the propagation of the electromagnetic field. Similar averaged 
expressions can be established for the center of the magnetic field 
intensity or the center of the electromagnetic field energy density. 
In the next section, this property of the group velocity is exploited 
to show the effect of the photonic crystal on the propagation of the 
electromagnetic field. 

\section{Computation of the dispersion law and applications\label{sec5}}

The dispersion law in photonic crystals has been investigated intensively with different
numerical methods. Since the eigenvalue problem (\ref{Shr}) is defined from 
the periodic operator $M(\x,\k)$ acting on the Hilbert space of square integrable periodic 
functions of $\x$, the most widely used numerical method was based on the expansion of 
the equations into the discrete Fourier (or plane-waves) basis \cite{LL90,ZS90,HCS90}. 
This expansion was used to predict photonic bandgap edges
\cite{HCS+94}, the effect of several structural imperfections on
such edges \cite{CN99}, the decay rate for single
photon emission in infinite structures \cite{LV01} and reflectivity 
and the inhibition of spontaneous emission for finite-thickness
structures \cite{Whi00}. 
An efficient software based on this method, MPB\footnote{https://mpb.readthedocs.io/} 
for \textit{MIT Photonic Bands}, has been elaborated by Steven Johnson and John 
Joannopoulos \cite{JJ01}. 

In the present chapter, the finite elements software comsol\cite{comsol} is used to solve 
eigenvalue problem (\ref{Shr}) for transverse electromagnetic waves allowed to propagate within an infinite periodic 
photonic crystal (in which case the computational domain reduces to the ``basic'' unit cell $\cell$ with Floquet-Bloch 
boundary conditions). The software comsol is also used to solve scattering (or forced) problems for transverse electromagnetic 
fields radiated by a line source placed within a finite photonic crystal (in which case some perfectly matched 
layers are required to ensure that the boundary of the computational domain is reflectionless).

\begin{figure}[ht]
\begin{center}
     {\includegraphics[width=70mm]{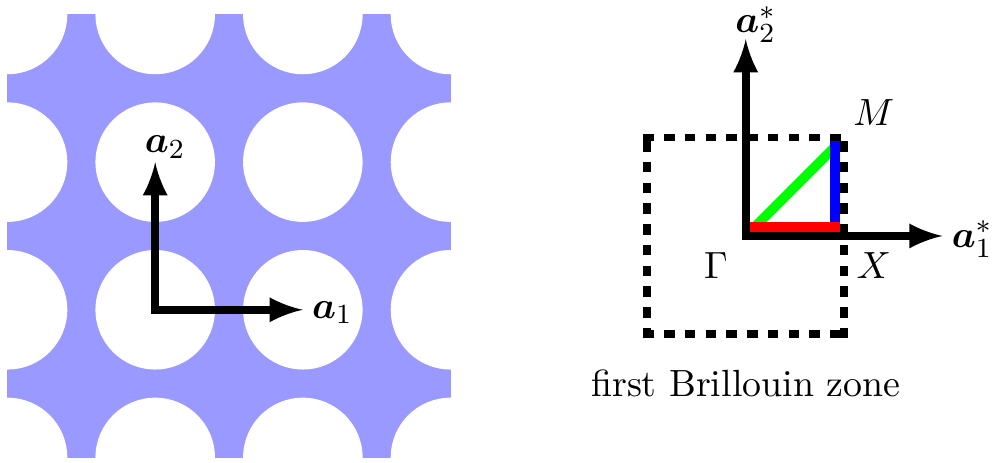}}
\end{center}
\caption{The considered two-dimensional photonic crystal made of circular holes drilled in a dielectric matrix. 
On the right the first Brillouin zone and the contour $\Gamma X M \Gamma$ of the reduced first 
Brillouin zone.
\label{figPC}}
\end{figure}

As an illustrative example, we consider a square photonic crystal with lattice constant $a$ and with 
circular air holes drilled in a matrix of permittivity $\ep_m = 11.29$ (refractive index $n=3.36$). This 
permittivity value corresponds to the effective index of $s$-polarized mode in a 
planar waveguide made of InGaAsP \cite{Fani2013}. The radius of the circular holes 
is set to $0.455a$, and this corresponds to the air filling ratio of $0.65$. The unit cell 
of the crystal is defined by the two vectors $\a_1 = a \, \e_1$ and $\a_2 = a \, \e_2$, and 
the first Brillouin zone by the vectors $\K_1 = (2 \pi / a) \, \e_1$ and 
$\K_2 = (2 \pi / a) \, \e_2$ (see Fig. \ref{figPC}). According to the symmetries of the unit cell, 
the dispersion law is represented on the path $\Gamma X M \Gamma$ in the first Brillouin zone with 
$\Gamma = (0,0)$, $X = (1/2,0)$ and $M=(1/2,1/2)$ in the basis $(\K_1,\K_2)$.
We compute the band diagrams associated with transverse electromagnetic waves propagating in the plane $(\a_1,\a_2)$. 
In this cylindrical case, one can split the two-dimensional spectral problem (\ref{thMe}) into the two scalar 
situations identified by  the $s$- and $p$-polarizations (also referenced by, respectively, the TM and TE cases \cite{JMW95}). Denoting 
by $E_{\#} \equiv E_{\#,3}$ and $H_{\#} \equiv H_{\#,3}$ the components of $\E_{\#}$ and $\H_{\#}$  along the axis $\e_3$ of invariance,
we look for pairs of eigenfrequencies and associated eigenfields, $(\omega,E_{\#})$ for $s$-polarization 
and $(\omega,H_{\#})$ for $p$-polarization, solutions of
\begin{equation}
\begin{array}{l}
- \partial_{\x} \cdot  \partial_{\x} E_{\#}(\x,\k,\om) = \om^2 \mu_0 \, \varepsilon(\x) \, E_{\#}(\x,\k,\om) \, , \\[2mm] 
-\partial_{\x} \cdot  \ep^{-1}(\x) \, \partial_{\x}  H_{\#}(\x,\k,\om) = \om^2 \mu_0 \, H_{\#}(\x,\k,\om) \, , 
\end{array}
\label{transmax}
\end{equation}
and such that $E_{\#}$ and $H_{\#}$ satisfy the Floquet-Bloch conditions (\ref{Bloch}) on the opposite edges 
of the periodic unit cell $\cell$.
Notice that, in the present two-dimensional case, the variables $\x$ and $\k$ are the two-components 
vectors $(x_1,x_2)$ and $(k_1,k_2)$. This problem is discretized using finite elements, and it is 
implemented with comsol\cite{comsol}.

\begin{figure}[ht]
\begin{center}
\hspace*{7mm}$s$-polarization\hspace*{43mm} $p$-polarization\\[1mm]
     \rotatebox{90}{\footnotesize \hspace*{5mm} normalized frequency \hspace*{1mm} $\om a / (2 \pi c)$}
     \hspace*{0.5mm}
     {\includegraphics[width=55mm,height=50mm]{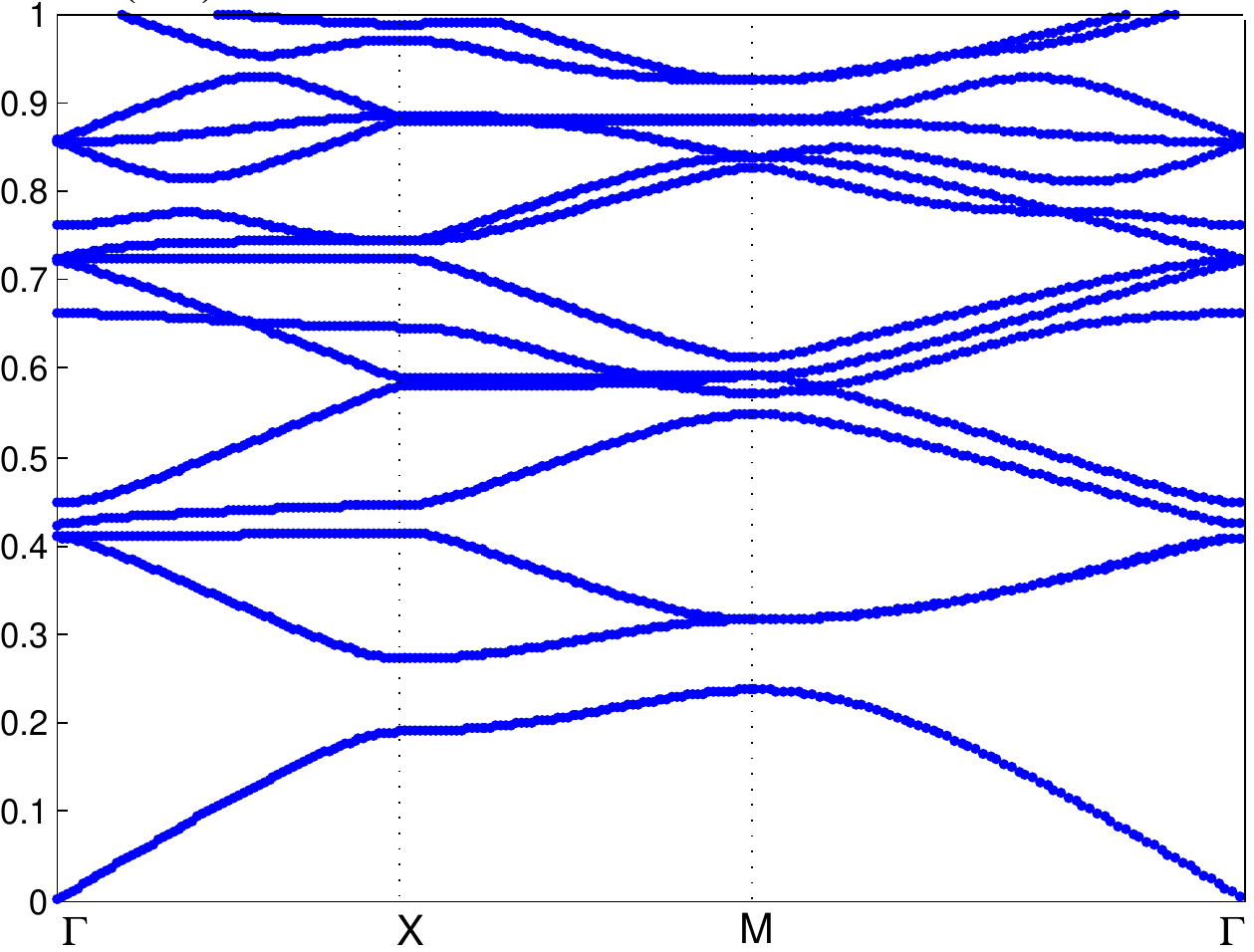}}
     \hspace*{3mm}
     \rotatebox{90}{\footnotesize \hspace*{5mm} normalized frequency\hspace*{1mm} $\om a / (2 \pi c)$}
     \hspace*{0.5mm}
     {\includegraphics[width=55mm,height=50mm]{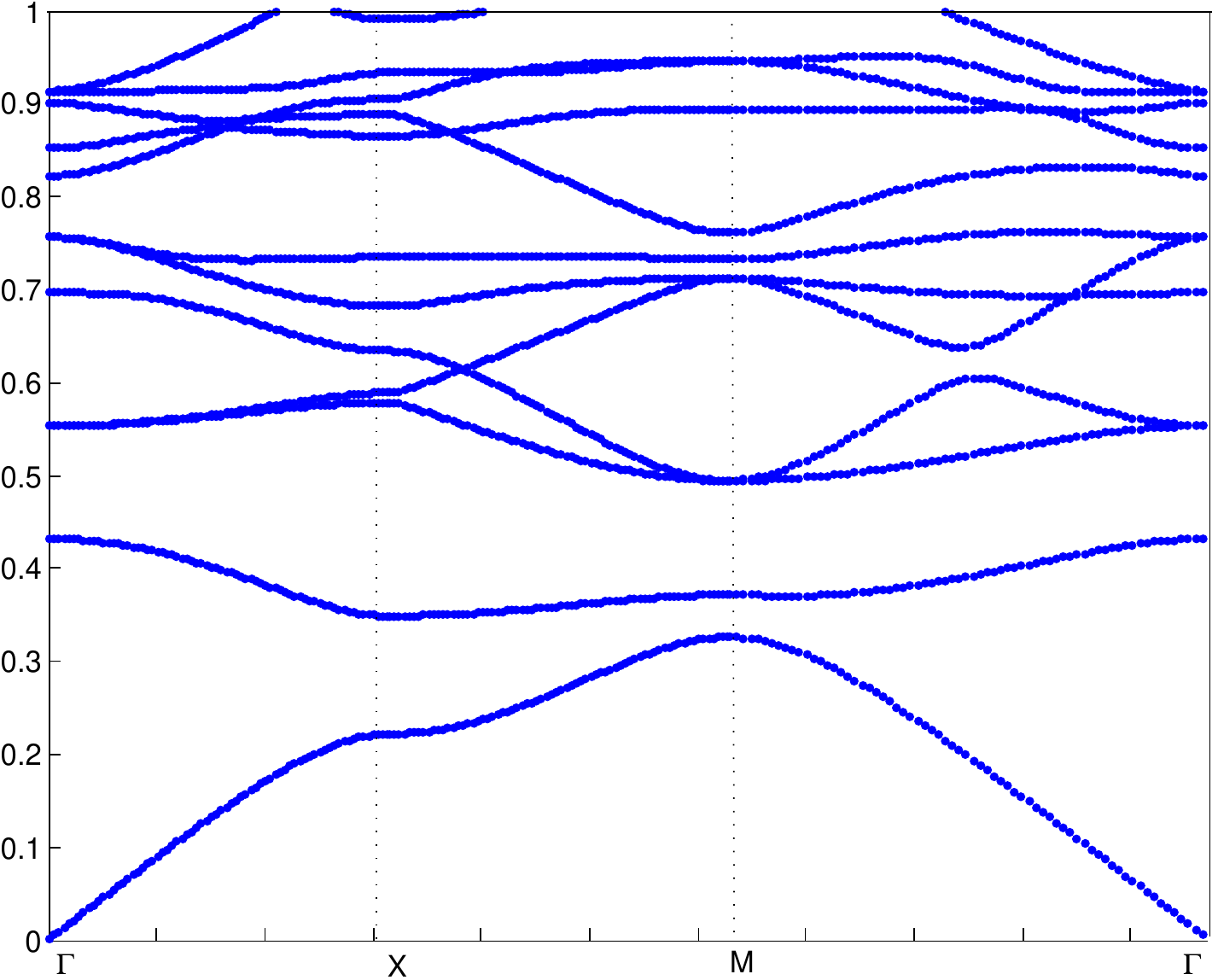}}
\end{center}
\caption{Dispersion diagrams in $s$- and $p$-polarizations: normalized frequency $\omega a/(2\pi c)$ versus 
Bloch wavevector $\k$ describing the first Brillouin zone contour $\Gamma XM \Gamma$ for transverse 
electromagnetic waves propagating within a doubly periodic square array of air holes of radius $0.455a$ 
of center-to-center spacing $a$ in a homogeneous isotropic medium of refractive index $n=3.36$.
\label{figDD}}
\end{figure}

Figure \ref{figDD} shows the dispersion diagrams in $s$- and $p$-polarizations as the normalized frequency 
$\om a / (2 \pi c)$ versus the Bloch wavevector describing the contour $\Gamma X M \Gamma$, 
$c = 1 / \sqrt{\ep_0 \mu_0}$ being the light velocity in vacuum. These diagrams report that two-dimensional 
photonic bandgaps exist around $\om a / (2 \pi c) \approx 0.26$ in $s$-polarization and  around 
$\om a / (2 \pi c) \approx 0.33$ or $0.48$ in $p$-polarization. Thus the polarized electromagnetic
fields cannot propagate at these frequencies. 
In addition to the bandgaps, the richness of the photonic bands $\om(\k)$
can be exploited in order to finely control the propagation of the electromagnetic field. 
The next part of this section is focused on the effect of effective anisotropy applied to source directivity.

A striking effect of directive emission of a field radiated by a line (respectively dipole line) source 
placed within a finite photonic crystal can be achieved at the frequency corresponding to an inflection point along the $XM$ 
direction of the lowest dispersion curve (known as acoustic band) in $s$- (respectively $p$-) polarizations.
Such inflection points appear along the $XM$ direction on the first band at normalized angular frequency $0.215$ 
in left panel of Fig. \ref{figDD} and at normalized frequency $0.26$ in right panel of Fig. \ref{figDD}. 
It can be expected at these inflection points that the dispersion law restricted to the corresponding 
frequencies displays vanishing curvature. Indeed, these points correspond to frontier between the situation of 
isofrequency contours of increasing size centered about $\Gamma$ and the situation 
of isofrequency contours of decreasing size centered about $M$. 

\begin{figure}[ht]
\begin{center}
\hspace*{10mm}$\big| E_3(\x,\om) \big|$ \hspace*{45mm} \mbox{$\big| H_3(\x,\om) \big|$\hspace*{11mm}}\\
\hspace*{0mm}\includegraphics[width=11cm]{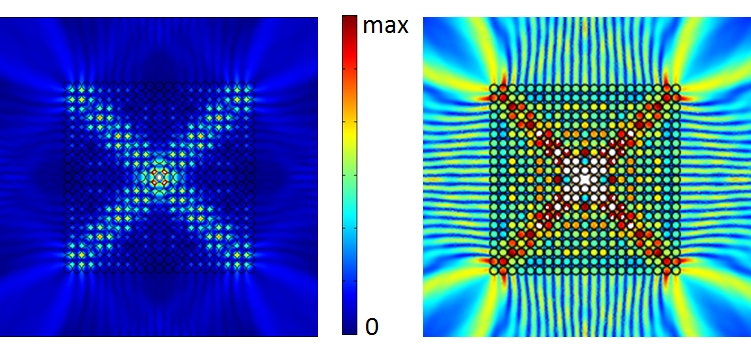}
\vspace*{-5mm}
\end{center}
\caption{Effect of effective anisotropy at inflection points. Left: in $s$-polarization for a line source 
at normalized frequency $\omega a/(2\pi c)=0.215$ placed at the center of the finite photonic crystal.
Right: in $p$-polarization for a dipole line source at normalized frequency $\omega a/(2\pi c)=0.26$ placed 
at the center of a finite photonic crystal. Note that $E_3$ and $H_3$ fields are radiated mostly along the $x=y$ and $x=-y$ directions.\label{figDA}}
\end{figure}

To illustrate this effect, we consider a line source in $s$-polarization (respectively dipole line source 
in $p$-polarization) placed in the center of a finite photonic crystal with 440 air holes. When we pickup 
a frequency close to the inflection point on the first band in left panel of Fig. \ref{figDD} 
(respectively in right panel of Fig. \ref{figDD}), we observe a striking effect in Fig. \ref{figDA} whereby 
light emitted by the source propagates preferably along the main diagonals of the PC. 
Notice that this behavior can be well predicted by the so-called high-frequency 
homogenization theory (HFH) \cite{Craster2010}. At these inflection points, the photonic crystal 
behaves like an effective medium described by an anisotropic refractive index with eigenvalues of 
opposite signs \cite{Antonakakis2013}. Note that the effective medium describing the photonic crystal in Fig. \ref{figDA} is actually isotropic in the low frequency regime, as the holes are circular, and achieving an anisotropy like in Fig. \ref{figDA} would require extremely elongated inclusions \cite{Guenneau2000}. 

\begin{figure}[ht]
\begin{center}
\hspace*{10mm}$\big| E_3(\x,\om) \big|$ \hspace*{45mm} \mbox{$\big| H_3(\x,\om) \big|$\hspace*{11mm}}\\
\hspace*{0mm}\includegraphics[width=11cm]{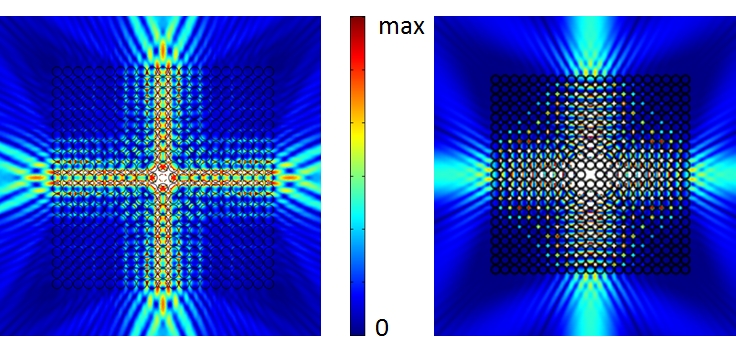}
\vspace*{-5mm}
\end{center}
\caption{Effect of extreme anisotropy at near flat bands. Left : $s$-polarization for a line source 
at normalized frequency $\omega a/(2\pi c)=0.4$ placed at the center of a finite photonic crystal.
Right : $p$-polarization for a dipole line source at normalized frequency $\omega a/(2\pi c)=0.35$ placed 
at the center of a finite photonic crystal.
White is outside color scale. Note that $E_3$ and $H_3$  fields are radiated mostly along the $x$ and $y$ directions.\label{figFlat}}
\end{figure}

Another interesting feature is the second band which is nearly flat for $p$-polarization 
in right panel of Fig. \ref{figDD}. If we consider the normalized frequency  $0.35$ for 
a dipole source placed in the center of the PC, we achieve a highly directive effect 
along the main horizontal and vertical axis in Fig. \ref{figFlat}. Likewise, the third 
band is flat along $\Gamma M$ for $s$-polarization in left panel of Fig. \ref{figDD},
and the highly directive source emission is also shown for a line source at the  
normalized frequency  $0.4$ in Fig. \ref{figFlat}. These infinitely anisotropic 
effective media in $s$and $p$ polarizations share some common features with ultra 
refractive optics \cite{EGT02,Enoch02}.   

\begin{figure}[b!]
\begin{center}
\hspace*{20mm}$\big| E_3(\x,\om) \big|$ \hspace*{50mm} \mbox{$\big| H_3(\x,\om) \big|$\hspace*{8mm}}\\
\hspace*{10mm}\includegraphics[width=11cm]{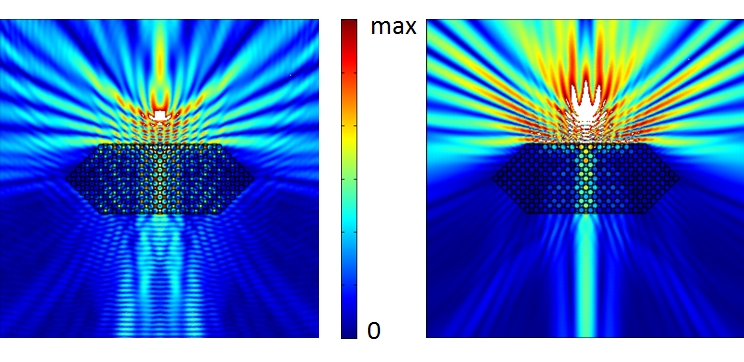}
\vspace*{-5mm}
\end{center}
\caption{Effect of effective anisotropy at inflection points in a rotated photonic 
crystal. Left: in $s$-polarization for a line source at normalized frequency 
$\omega a/(2\pi c)=0.215$ placed above a slice of the finite photonic crystal 
in Fig. \ref{figDA} rotated through an angle $pi/4$ about its center. Right: in 
$p$-polarization for a dipole line source at normalized frequency $\omega a/(2\pi c)=0.26$. 
Note that $E_3$ and $H_3$ fields are guided mostly along the $y$ direction.\label{figLens}}
\end{figure}

Finally, the vital role of the interface of the photonic crystal is exemplified in Fig. 
\ref{figLens}. The situation of $p$-polarization (respectively $p$-polarization) at 
normalized frequency $\omega a/(2\pi c)=0.215$ (respectively $\omega a/(2\pi c)=0.26$) 
is considered with the photonic crystal lattice rotated angle of $\pi/4$ about the 
center of the finite crystal. Two interfaces are sliced in the direction 
$(\a_1 + \a_2) / \sqrt{2}$. In that case, the wavevectors making a small angle with 
the normal to the interface correspond to the main diagonals of the photonic crystal 
in Fig. \ref{figDA}. As a result, a self-collimation along the $M\Gamma$ direction is 
achieved in both polarizations and a focusing effect can be observed in $s$-polarization, 
see Fig. \ref{figLens}. Note that one could implement an algorithm as described in 
\cite{Smigaj2012} in order to reduce the impedance mismatch (and thus improve the source 
coupling) between crystal and surrounding medium.

Most of the features such as high-directivity and lensing effects can be captured 
by the high frequency homogenization. The essence of this asymptotic method is that 
one introduces two separate scales ${\y}$( the macroscopic scale) and $\mathbf{\xi}$ 
(the microscopic scale) and then perturbs away from high-symmetry points in the 
Brillouin zone and there the ansatz $H_{\#}({\y},\mathbf{\xi},\k,\om)
=u_0({\y},\mathbf{\xi})+\delta u_1(\y,\mathbf{\xi})+\delta^2 u_2({\y}, 
\mathbf{\xi})+\ldots $ is posed for the field $H_{\#}$ in $p$-polarization 
(respectively $E_{\#}$ in $s$-polarization) and $\omega^2 =\Omega_0^2+\delta 
\Omega_1^2+\delta^2 \Omega_2^2+\ldots$ for the frequency $\om$ in (\ref{transmax}) 
where $\delta$ is a small positive parameter and we have dropped the implicit 
dependence on $\k$ and $\omega$ in $u_i's$ 
to lighten notations. The highly oscillatory functions $u_i's$ on short scale 
$\mathbf{\xi}$ get modulated by the slowly varying long scale ${\y}$.
Importantly, $\Omega_0$ is the frequency corresponding to a standing wave $u_0$ 
associated with a Bloch vector chosen at a high symmetry point of the Brillouin 
zone (i.e. $\Gamma$, $X$ or $M$). The $u_i({\y},\mathbf{\xi})$'s adopt the 
boundary conditions on the edge of the cell (so periodicity or anti-periodicity 
as we are at a high-symmetry point).  An ordered set of equations emerge indexed 
with their respective power of $\eta$, and are treated in turn. The leading order 
approximation (or homogenized field) $u_0$, and subsequently $u_{j}$, are computed 
using the standard finite element package comsol \cite{comsol}, although many other 
numerical methods could be used instead. It is then possible to replace the periodic 
structure by an effective medium described by an anisotropic tensor. For instance, 
comsol computations give the following effective tensor $T_{11}=-8.6656$, 
$T_{22}=0.9209$, $T_{12}=T_{21}=0$ when we perturb away from the $X$ symmetry point 
towards the inflection point on the second band in the right panel of Fig. \ref{figDD}. 
This extremely anisotropic tensor agrees well with the directive emission of the dipole 
source in Fig. \ref{figFlat}, see right panel. The theory of high frequency homogenization is introduced 
in \cite{Craster2010} and some numerical illustrations for photonic crystals given 
in \cite{Antonakakis2013}.

Thus far, we have studied the richness of the dispersion law in dielectric photonic crystals. However,  
the possibility offered by the metallic materials to obtain a photonic bandgap in the visible 
range \cite{Mor99,LM00} motivates use of augmented formalisms of Maxwell's equations \cite{Tip98,GT10} 
for absorptive media : Metals are inherently absorptive at optical wavelengths. 
This is the topic of the next sections.

\section{The dispersion law in dispersive and absorptive photonic crystals\label{sec6}}

The notion of dispersive and absorptive photonic crystal appeared with the possibility 
offered by the metallic materials to obtain a photonic bandgap in the visible 
range \cite{Mor99,LM00}. The investigations on dispersion and absorption 
led to the definition by Adriaan Tip of the \textbf{auxiliary field formalism} 
\cite{Tip97,Tip98}, which extends the Maxwell's equations to a classical evolution 
equation with a time-independent selfadjoint operator and thus allow{s} the simplification 
of dispersion and absorption. This formalism has been used to propose a definition of 
photonic bandgaps in dispersive and absorptive photonic crystals \cite{Tip00} and to 
show that spontaneous emission cannot be inhibited in presence of absorption \cite{Tip00}. 
Then, the computation of the dispersion law in dispersive and absorptive photonic crystals has 
been performed in the case of two-dimensional photonic crystals of circular \cite{Han03} 
and square \cite{Com03} rods made of a Drude metal. These preliminary calculations have 
been performed solving time-harmonic Maxwell's equations and using algorithms finding 
the complex eigenfrequencies $\om(\k)$ as roots of a linear system in the complex plane. 
A more efficient numerical method has been proposed in \cite{Fan10} by implementing in the 
numerical program a simplified version of the auxiliary field formalism. This method leads 
to an extension of Maxwell's equations with no dispersion but with remaining absorption. 
This method is now a crucial tool in the calculation of quasi-normal modes 
\cite{lalanne2018,Wei2018}. 

In this section, the example considered in \cite{Han03,Bru16} is revisited. 
The geometry is the same as in the previous section, except that the photonic 
crystal is made of circular rods of a Drude metal in air. 
The $s$-polarization is solely considered. The permittivity of the 
Drude metal is set to
\begin{equation}
\ep(\om) = \ep_0 - \ep_0 \, \dfrac{\om_p^2}{\om (\om+ i \gamma)}\, , 
\qquad \dfrac{\om_p a}{2 \pi c} = 1.1 \, , \qquad \dfrac{\gamma a}{2 \pi c} = 0.05 \, .
\label{epDrude}
\end{equation}
The computation of the eigenfrequencies is based on a version of the auxiliary field formalism 
that allows the linearization of the spectral problem associated with frequency dispersive 
materials described by Drude or Lorentz models \cite{Bru16}. Here, the auxiliary 
field 
\begin{equation}
\A(\x,t) = - 2 i \, \dfrac{\om_p}{\sqrt{2}} \displaystyle\int_{-\infty}^t ds \, 
\exp[- \gamma (t-s)] \, \E(\x,t-s)
\end{equation}
is added to the electromagnetic field ${\E}(\x,t)$ and $\H(\x,t)$ to express Maxwell’s 
equations as an augmented operator independent of time (see reference \cite{Bru16}). 
A variational form of the resulting augmented system is derived and {discretized using} 
the finite element method (FEM).

\begin{figure}[ht]
\begin{center}
     {\includegraphics[width=75mm]{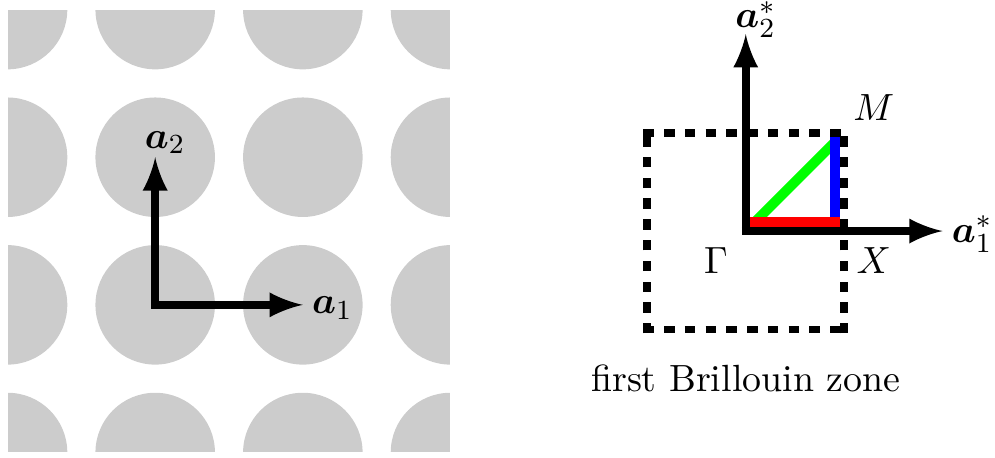}}
     \vspace*{-3mm}
\end{center}
\caption{The considered two-dimensional photonic crystal made of circular rods in air. 
On the right the first Brillouin zone and the contour $\Gamma X M \Gamma$ of the reduced first 
Brillouin zone: $\Gamma X$ is the red line, $X M$ is the blue line and $M \Gamma$ is the green line.
\label{figDrude}}
\end{figure}

{One cell of the structure} described in Fig.~\ref{figDrude} is meshed using 
the GNU software Gmsh \cite{gmsh}. First or second order edge elements (or Webb 
elements\cite{webb1993hierarchal}) are used. The GetDP \cite{getdp} {GNU} software allows
to handle the required various basis functions handily. Very recent progress in sparse matrix 
eigenvalue solvers allow to tackle the discrete problem very efficiently. For the purpose of this 
study, we interfaced GetDP with two particularly well suited and recent solvers of the SLEPc 
library \cite{Hernandez:2005:SSF} dedicated to solve large scale sparse eigenvalue problems.

The usual representation of the dispersion law in periodic structures provides 
the eigenfrequencies $\om(\k)$ for Bloch wavevector $\k$ describing the contour 
of the reduced first Brillouin zone. This representation is suitable for real 
eigenfrequencies since it maps the one-dimensional contour $\Gamma X M \Gamma$ 
to the real axis of frequencies $\om$. In the present case with absorption, the 
wavevectors $\k$ are mapped to complex eigenfrequencies. Hence it is relevant 
to describe the whole (two-dimensional) surface of the first reduced Brillouin zone 
and to obtain the corresponding surfaces in the complex plane of frequencies. 

\begin{figure}[h!]
     \vspace*{2mm}
\begin{center}
     {\includegraphics[width=0.80\textwidth]{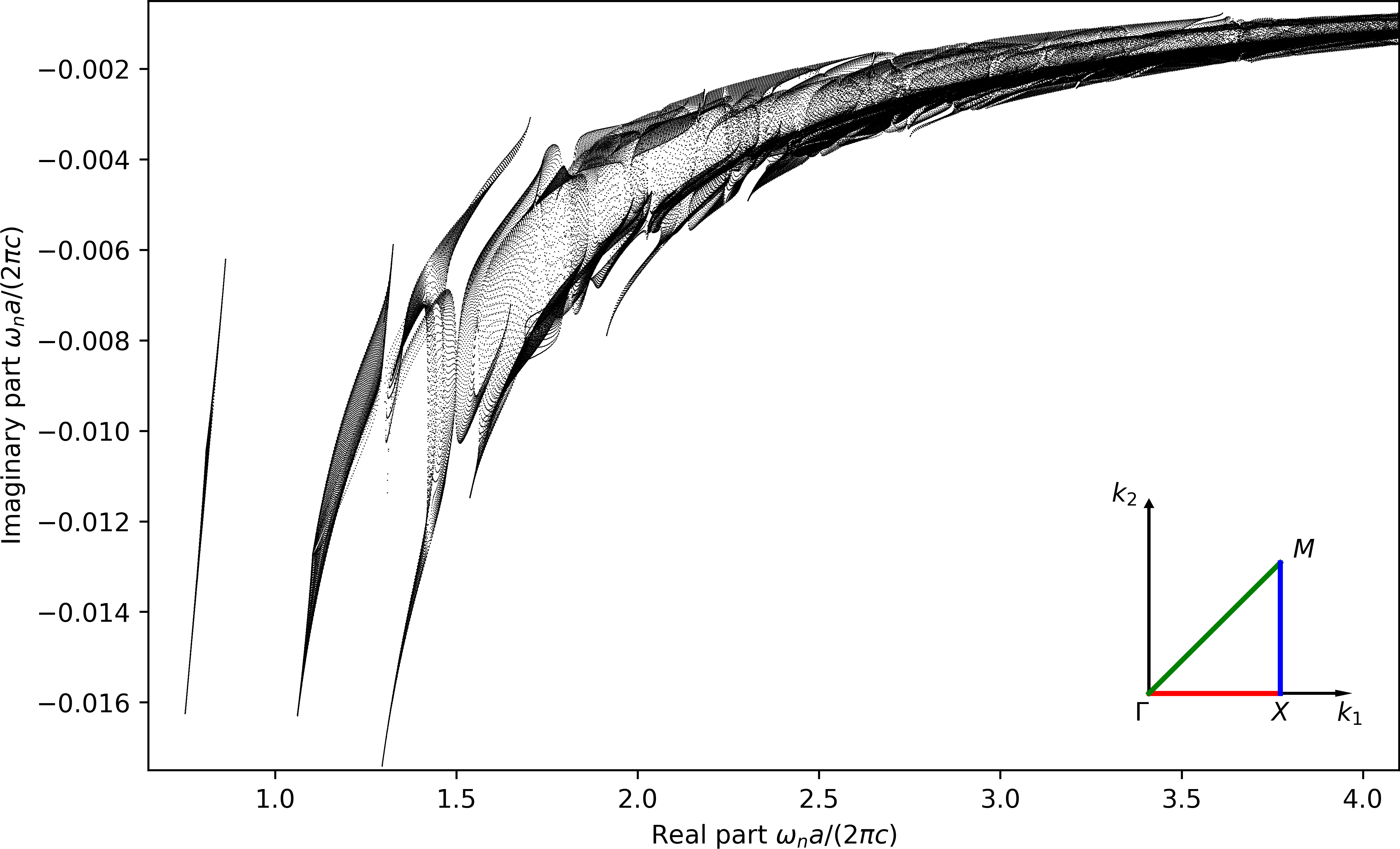}}
     \vspace*{-4mm}
\end{center}
\caption{Complex spectrum of the normalized eigenfrequencies$\om_n a / (2\pi c)$ of
photonic crystals made of and circular rods for all the wavevectors $\k$ 
in the first reduced Brillouin zone.\label{s-spectum}}
\end{figure}

Figure \ref{s-spectum} shows for $s$-polarization the whole spectrum of resonances $\om(\k)$ 
in the photonic crystal made of the Drude circular rods. The metallic nature of the Drude 
material around 
the null frequency leads to the absence of photonic bands in the range of low frequencies. 
Then, the first band appears well-separated of the remaining spectrum by a band around 
$\om a / (2 \pi c) = 1.0$, which is associated to a true photonic bandgap when the 
absorption parameter $\gamma$ is set to zero. The next bands appear to overlap in 
the complex plane to finally merge at the high frequency with the real axis. 
Hence the set of resonances tends toward the spectrum of the free Laplacian at high 
frequencies which is consistent with the behavior $\ep(\om) \rightarrow \ep_0$ of the 
Drude permittivity (\ref{epDrude}) when $| \om | \rightarrow \infty$. 


\begin{figure}[h]
     \vspace*{-2mm}
\begin{center}
     {\includegraphics[width=0.95\textwidth]{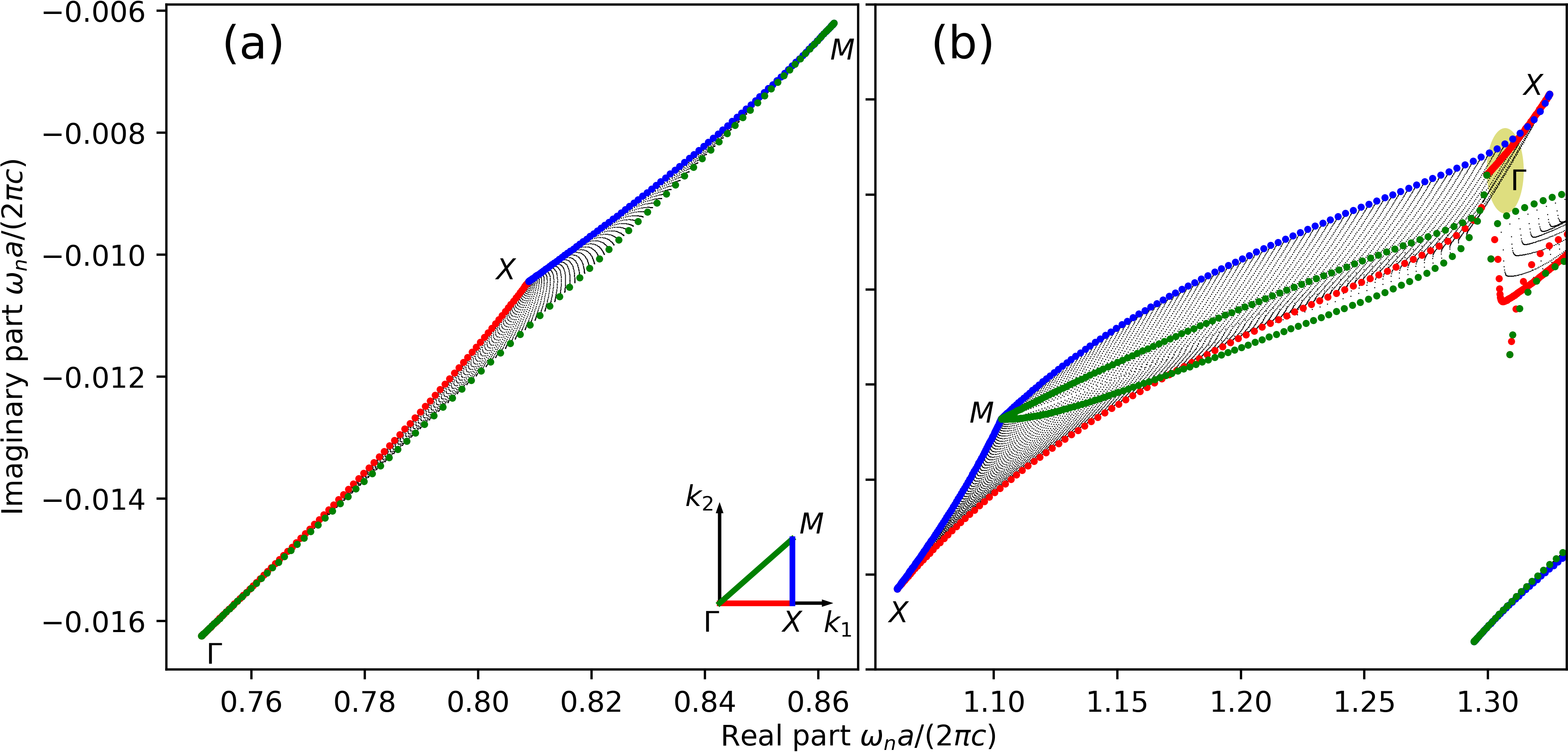}}
     \vspace*{-4mm}
\end{center}
\caption{First band (a) and second and third bands (b) of the complex spectrum of the 
normalized eigenfrequencies $\om_n a / (2 \pi c)$.\label{bands123}}
\end{figure}

\begin{figure}[h!]
\begin{center}
     {\includegraphics[width=0.95\textwidth]{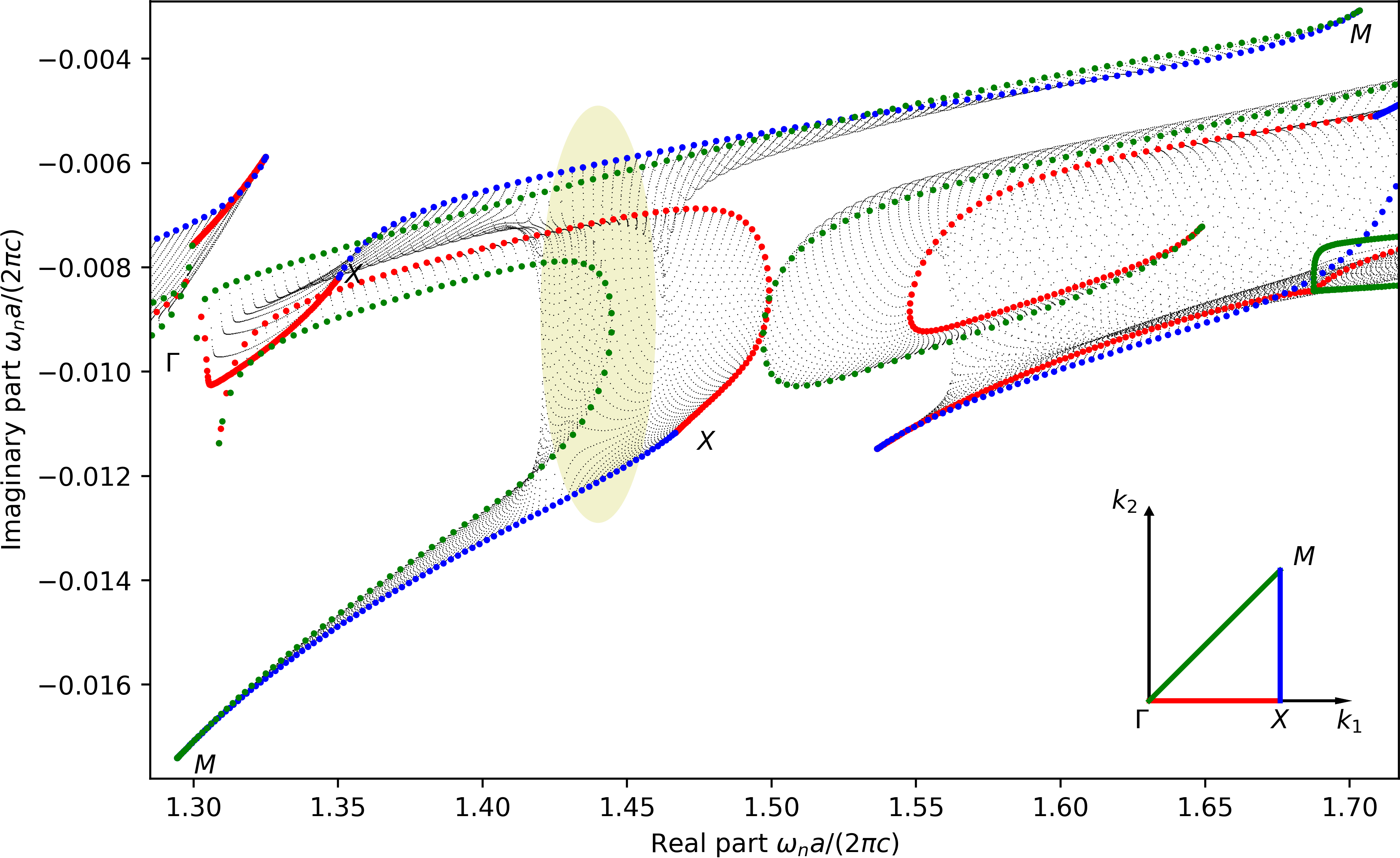}}
     \vspace*{-4mm}
\end{center}
\caption{Bands 4, 5 and of higher order of the complex spectrum of the 
normalized eigenfrequencies $\om_n a / (2 \pi c)$.\label{bands45}}
\end{figure}

Figure \ref{bands123} shows the first band (left panel) and the overlapping 
second and third bands (right panel). The first band is a skewed triangle which 
can be unambiguously related to the reduced first Brillouin zone. Each edge of this 
triangle $\Gamma X M \Gamma$ can be clearly identified: $\Gamma X$ is the red line, 
$X M$ is the blue line and $M \Gamma$ is the green line. Second and third bands overlap 
but can still be well-identified. However, one can remark that some resonances 
around the yellow annotation in Fig. \ref{bands123} seem to {lie} outside the skewed 
triangles. Such a phenomenon can be clearly observed on Fig. \ref{bands45} where, for instance, 
resonances are located between the bands 4 and 5 and seem to produce a connexion between 
these bands around the yellow annotation. This phenomenon, already pointed out in 
the literature \cite{Han03,Bru16}, is a counterexample of a {widely accepted assumption}
in the case of non absorptive and non dispersive periodic structures: \textit{`` the 
eigenfrequencies corresponding to the contour $\Gamma X M \Gamma$ are the extrema of 
eigenfrequencies of Bloch modes''}. According to this assumption, all the 
eigenfrequencies should lie inside the skewed triangle corresponding to 
$\Gamma X M \Gamma$. This counterexample shows the necessity to describe the whole 
inside of the reduced Brillouin zone to get a complete picture of the Bloch spectrum 
in the case of dispersive and absorptive photonic crystals.

The Bloch spectrum of dispersive and absorptive photonic crystals reveals a rich structure 
with resonances outside the contour of the reduced first Brillouin zone. As shown in 
reference \cite{Bru16}, this structure of the spectrum shows in addition cups and loops 
with potential effects on the group velocity. These first investigations, which start from 
the auxiliary field formalism, represent preliminary results on the vast field of 
spectral theory of non-selfadjoint operators. 

\section{Analytic properties of the dispersion law\label{sec7}}

The dispersion law $\om(\k)$ in photonic crystal is usually considered for 
real Bloch wavevector $\k$ and frequency $\om$. In the case of absorptive 
photonic crystal, it has been seen in the previous section that, for 
real wavevector $\k$, the corresponding frequencies $\om(\k)$ take complex 
values with negative imaginary part. In these two cases, the curves $\om(\k)$ 
have no global analytic structure since they are made of 
bands separated by gaps. However, the definition (\ref{Shr}) of the operator 
$M(\x,\k)$ can be extended to complex wavevectors $\k$. Thus its eigenvalues $\om(\k)$ 
and the dispersion law, which can be \textit{formally} defined by the equation 
\begin{equation}
\text{det} \, \big| M(\x,\k) - \om \big| = 0 \, ,
\label{det}
\end{equation}
can be extended to complex wavevector $\k$. Conversely, the equation above can be 
extended to complex frequencies $\om$ and the dispersion law is then defined by the 
roots of (\ref{det}) as $\k(\om)$. 

The aim of this section is to provide arguments supporting that the dispersion law 
$\k(\om)$ is an analytic function of the complex frequency $\om$ in the domain 
of positive imaginary parts Im$(\om) > 0$. Conversely, it is reasonable to consider 
that the dispersion law 
$\om(\k)$ is an analytic function of $\k$ as soon as Im$(\k) \neq 0$. 
The idea proposed in this section is to use the physical argument stating that the 
electromagnetic field cannot propagate faster than $c = 1 / \sqrt{\ep_0 \mu_0}$, 
the light velocity in vacuum.
Thus, the electromagnetic field should be strictly included in a domain defined 
by a condition similar to $| \x | < c t$. Then, a Paley-Wiener argument \cite{RSII} could 
be used to show the equivalence between the compact support of the electromagnetic field 
related to $| \x | < c t$ and the analytic properties with respect to the variables 
$\k$ and $\om$ in the corresponding Fourier space. These arguments will be developed 
in the next part of this section. Finally, it is mentioned that a rigorous derivation 
in the two-dimensional case can be found in reference \cite{Knorrer1990}. \\

\noindent
\textbf{The analytic properties of the solution of Maxwell's equations.} 
The time-dependent Maxwell equations (\ref{Me}) are considered with the current source 
density $\J(\x,t)$ switched on at the initial time $t = 0 $ and homogeneous in the 
ball of radius $c  t_0 > 0$ and centered at the origin $\x = 0$:
\begin{equation}
\J(\x,t) = \J_{\!0} \, \theta(c t_0 - |\x|) \, \theta(t) \sin[\om_s t] \, , 
\label{sourceJ}
\end{equation}
where $\J_{\!0}$ is a constant vector in $\R^3$, 
$\om_s$ is the oscillating frequency of the source and 
$\theta(t)$ is the step function: $\theta(t) = 0$ for $t<0$ and $\theta(t) = 1$ for $t\ge 0\,$. 
Let $G(\x,t)$ be the electric field radiated by this electromagnetic source : 
from Maxwell's equations (\ref{Me}),
 \begin{equation}
\mu_0 \, \ep(\x) \, \partial_t^2 G(\x,t)  + \rot \rot G(\x,t) = - \mu_0 \,\J_{\!0} \, 
\theta(c t_0 - |\x|) \, \theta(t) \, \om_s \cos[\om_s t]  \, 
\label{evolution}
\end{equation}
According to the causality principle, the field $G(\x,t)$ radiated by this source 
switched on at $t=0$ must vanish for all negative times. Moreover, since 
the electromagnetic field cannot propagate faster than the light velocity 
in vacuum $c = 1 / \sqrt{\ep_0 \mu_0}$, we must have for $t > 0$
\begin{equation}
c \, (t + t_0) < | \x | \quad \Longrightarrow \quad G(\x,t) = 0 \, . 
\label{cone}
\end{equation}
First, the analytic property with respect to the complex frequency $\om$ is derived. 
Since the function $G(\x,t)$ vanishes for all times $t < 0$, the Laplace transform 
\begin{equation}
\widetilde{G}(\x,\om) = \displaystyle\int_{0}^\infty dt \, \exp[i \om t] \, G(\x,t)
\label{Laplace}
\end{equation}
is well-defined for all complex frequency with positive imaginary part Im$(\om)$. 
Indeed, the field $G(\x,t)$ cannot increase faster than linearly with time
\footnote{Strictly speaking, the electromagnetic energy cannot increase faster than 
linearly with time since the equation (\ref{evolution}) can be written as an evolution 
equation (satisfying the Hille-Yosida theorem) which involves a dissipative operator with an 
excitation whose energy is uniformly bounded in time \cite{Dau-00}. In practice, it is reasonable 
to consider that it 
remains true for $G(\x,t)$ in standard physical situations, in particular with 
the source (\ref{sourceJ}).}, 
while the exponential factor $\exp[i \om t]$ introduces the exponential decrease 
$\exp[ - \text{Im}(\om) \, t]$. 
The integral (\ref{Laplace}) as well as all its derivatives with respect to 
the frequency: 
\begin{equation}
\dfrac{d^p \widetilde{G}}{d \om^p}(\x,\om) = \displaystyle\int_{0}^\infty dt \, (i \om)^p \, 
\exp[i \om t] \, G(\x,t) \, , \quad \forall \,  p \, \in \, \mathbb{N} \, , 
\label{dGtilde}
\end{equation}
are well-defined since the integrated function are integrable for $\om$ in the upper half-plane.
Thus the Laplace transform $\widetilde{G}(\x,\om)$ is analytic with respect to the complex
frequency $\om$ in the domain Im$(\om) > 0$. The initial field field $G(\x,t)$ can be 
retrieved using the inverse Laplace transform:
\begin{equation}
G(\x,t) = \dfrac{1}{2\pi} \, \displaystyle\int_{R_\eta} dt \, \exp[ - i \om t] \, 
\widetilde{G}(\x,\om) \, ,
\label{invLaplace}
\end{equation}
where $R_\eta$ is the line parallel to the real axis of complex frequencies 
with the imaginary part set to $\eta > 0$, 
\begin{equation}
R_\eta = \big\{ \omega \in \C \, | \, \text{Im}(\om) = \eta \big\} \, .
\end{equation}
For negative times $t$, the integral (\ref{invLaplace}) can be calculated by closing the 
contour integral with a semi circle in the upper half plane of complex frequencies. 
Therefore, the analytic property of the Laplace transformed $\widetilde{G}(\x,\om)$ in the 
domain Im$(\om) > 0$ implies that $G(\x,t) =0$ for $t < 0$. Hence the following 
equivalence has been established:
\begin{equation}
G(\x,t) = 0 \quad \forall \, t < 0 \quad \Longleftrightarrow \quad \widetilde{G}(\x,\om) \quad  \text{analytic} 
\quad \forall \, \om \quad \text{with} \quad \text{Im}(\om) > 0 \, .
\label{equiv-om}
\end{equation}
Second, the analytic property with respect to the complex vector $\k$ is investigated. 
Assuming first a Fourier transform  of the field with respect to the space variable $\x$, 
and then a Laplace transform (\ref{Laplace}), yields
\begin{equation}
\begin{array}{ll}
\dbtilde{G}(\k,\om) & = \displaystyle\int_{0}^\infty dt \, \exp[i \om t] \, 
\displaystyle\int_{\R^3} d\x \, \exp[- i \k \cdot \x] \, G(\x,t) \\[4mm]
& = \displaystyle\int_{0}^\infty dt \, 
\displaystyle\int_{|\x| < c(t+t_0)} d\x \, \exp[i (\om t - \k \cdot \x)] \, G(\x,t) \, ,
\label{Fourier}
\end{array}
\end{equation}
where the causality condition (\ref{cone}) has been used. For frequency $\om$ with positive 
imaginary part, the exponential function under the integral can be bounded by 
\begin{equation}
\begin{array}{ll}
\big| \exp[i (\om t - \k \cdot \x)] \big| & \leq \exp[ - \text{Im}(\om) t + |\text{Im}(\k) \cdot \x | ] \\[2mm]
& \leq \exp[ - \text{Im}(\om) t + |\text{Im}(\k) | \, c (t + t_0) ]\\[2mm]
& \leq \exp[ - \{ \, \text{Im}(\om) - c \, |\text{Im}(\k)| \, \} \, t \, ] \: \exp[ \, c \, |\text{Im}(\k)| \, t_0 \, ] \, .
\end{array}
\label{bound}
\end{equation}
Thus the integral expression (\ref{Fourier}) of $\dbtilde{G}(\k,\om)$ is well-defined if 
$\text{Im}(\om) - c \, |\text{Im}(\k)| > 0$, and that remains true for all the derivatives 
of $\dbtilde{G}(\k,\om)$ with respect to $\om$ and to (the components of) $\k$. 
The space-time extension of (\ref{equiv-om}) is then
\begin{equation}
G(\x,t) = 0 \quad \text{if} \quad c \, (t + t_0) < | \x |  \quad 
\Longleftrightarrow \quad \dbtilde{G}(\k,\om)  \quad  \text{analytic if} 
\quad \text{Im}(\om) > c \, |\text{Im}(\k)| \, .
\label{equivalence}
\end{equation}
This relationship shows the equivalence between the causality principle and the analytic properties in 
the Fourier--Laplace space. Next, the objective is to transfer some analytic properties from the 
solution $\dbtilde{G}(\k,\om) $ to the dispersion law $\om(\k)$. \\

\noindent
\textbf{The analytic properties of the dispersion law.}
The following section gives a reasoning that enlightens the analytic properties 
of the dispersion law with respect to frequency in the upper-half plane
but it is not strictly speaking a rigorous mathematical proof. 
To our knowledge, showing the analytic regularity of the dispersion law is still 
an open problem in mathematics for three-dimensional photonic crystals, 
but it has been established for the two-dimensional case in \cite{Knorrer1990}.

The solution $G(\x,t)$ of Maxwell's 
equations can be retrieved from the inverse Fourier-Laplace transform applied to 
(\ref{Fourier}):
\begin{equation}
G(\x,t) = \dfrac{1}{(2 \pi)^4} 
\displaystyle\int_{R_\eta} d\om \, \displaystyle\int_{\R^3} d\k \, 
\exp[ - i \om t] \, \exp[i \k \cdot \x] \, \dbtilde{G}(\k,\om) \, .
\label{invFourier}
\end{equation}
Let $\ks = |\k| = \sqrt{\k \cdot \k}$ be the modulus of $\k$ 
and $\e_\k = \k / \ks$ 
the unit vector pointing in the direction of $\k$. 
The integral over the wavevectors $\k$ is performed with respect to the 
wavenumber $\ks$ and their directions $\e_\k$ on the unit sphere $S$: 
\begin{equation}
G(\x,t) = \dfrac{1}{(2 \pi)^4} 
\displaystyle\int_{R_\eta} \!\! d\om \, \displaystyle\int_{S} d\e_\k \displaystyle\int_{0}^{\infty} 
\! d\ks \,  \ks^2\, 
\exp[ - i \om t] \, \exp[ i \ks \, \e_\k \cdot \x \, ] \, \dbtilde{G}(\k,\om) \, .
\label{Gxt}
\end{equation}
Let $S_\x^+$ be the hemisphere defined by 
\begin{equation}
S_\x^+ = \big\{ \e_\k \in S \, | \, \e_\k \cdot \x > 0 \, \big\} \, . 
\label{Splus}
\end{equation}
Then, the expression (\ref{Gxt}) can be written 
\begin{equation}
G(\x,t) = \dfrac{1}{(2 \pi)^4} 
\displaystyle\int_{R_\eta} \!\! d\om \, \exp[ - i \om t] 
\displaystyle\int_{S_\x^+} d\e_\k \displaystyle\int_{\R} 
\! d\ks \,  \ks^2\, \exp[ i \ks  \, \e_\k \cdot \x \, ] \, \dbtilde{G}(\k,\om) \, .
\label{Gxtspherique}
\end{equation}
\begin{figure}[ht]
\begin{center}
     {\includegraphics[width=125mm]{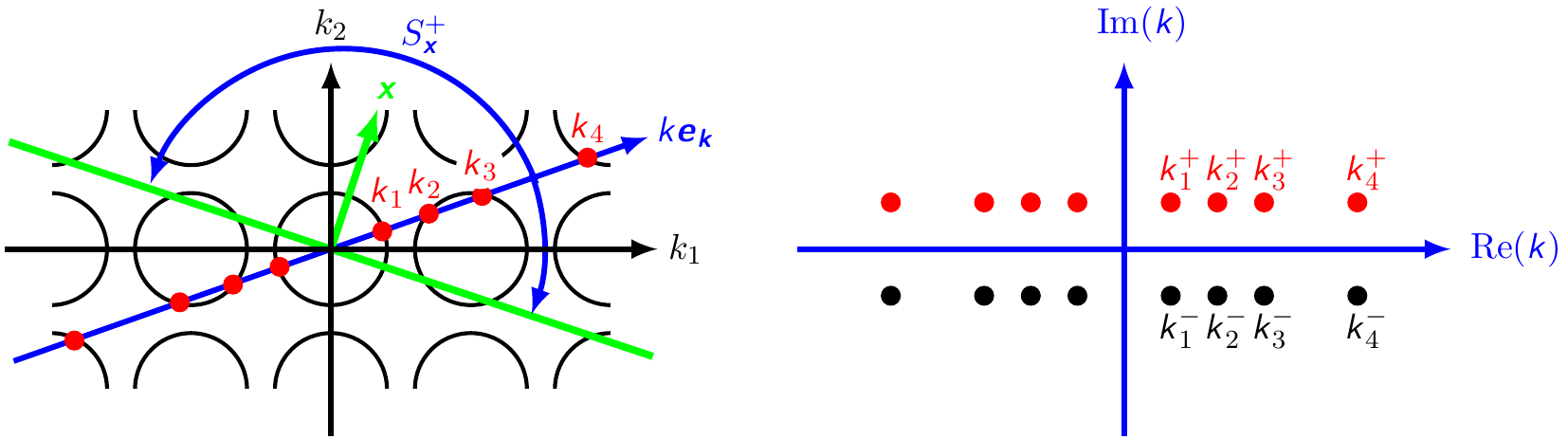}}
\end{center}
\caption{A representation of the poles $\ks_p^+ = \ks_p^+(\om,\e_\k)$. 
Left: the dispersion law for a real frequency represented as periodic circles. 
For the direction $\e_\k$, the wavevector $\ks \e_\k$ meets a discrete infinite 
number of Bloch wavevectors. When the direction $\e_\k$ describes all the hemisphere 
$S_\x^+$, the whole dispersion law made of the circles periodically 
arranged are crossed by the line $\ks \e_\k$. Right: for a small positive 
imaginary part, the poles $\ks_p$ becomes the pair of poles $\ks_p^\pm$ with 
positive or negative imaginary parts. The description of $S_\x^+$ by the 
direction $\e_\k$ leads to the selection of the poles with positive 
imaginary part.\label{figDL}
}
\end{figure}

\begin{figure}[ht]
\begin{center}
     {\includegraphics[width=125mm]{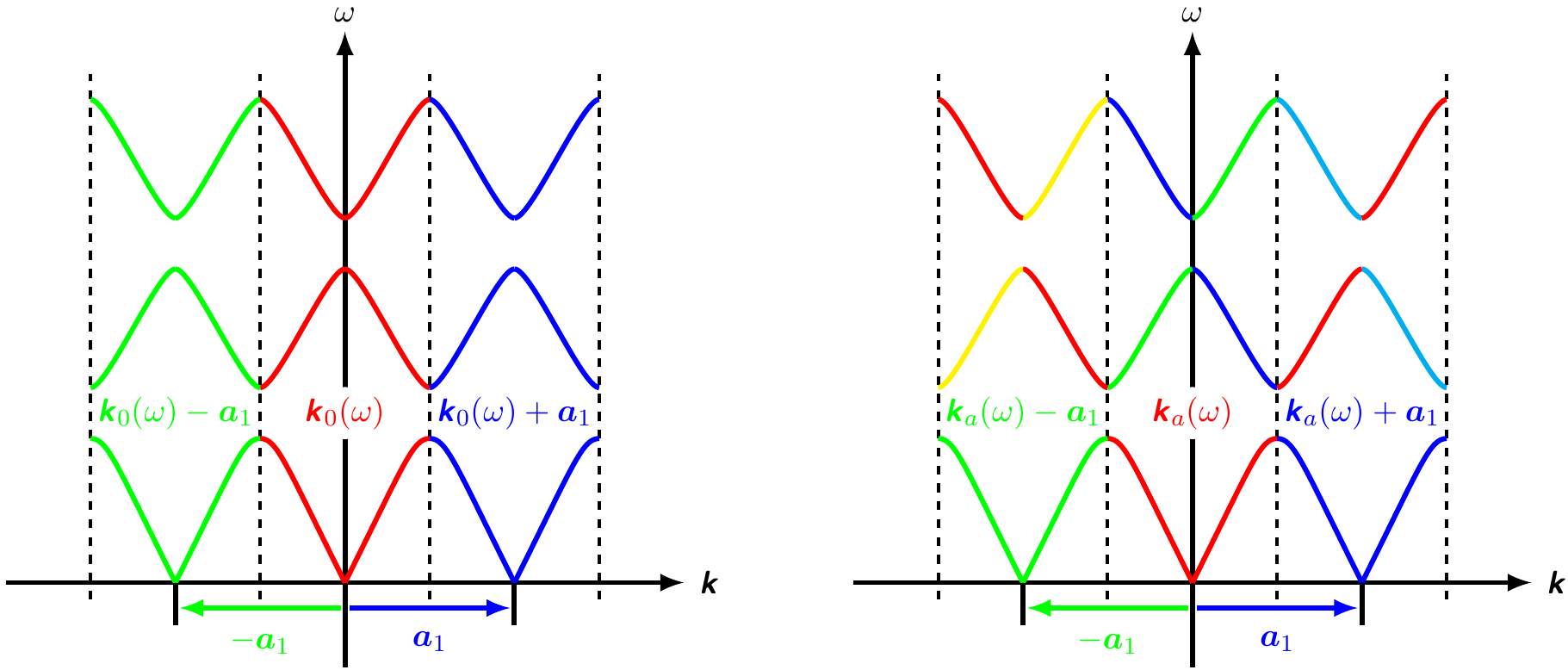}}
\end{center}
\caption{A representation of the periodic dispersion law for real frequencies 
as the function $\k(\om)$. 
Left: the folded dispersion law $\k_0(\om)$ restricted to the First Brillouin zone is represented in red. 
This curve presents a lack of analyticity at the boundaries of the first Brillouin zone. The whole 
periodic dispersion law is obtain by translating $\k_0(\om)$ by all the vectors $\a$ in the lattice $\lat$. 
The folded parts $\k_0(\om) + \a_1$ and $\k_0(\om) - \a_1$ are respectively represented in blue and green. 
Right: the unfolded dispersion law $\k_a(\om)$ is drawn in red and the parts $\k_a(\om) + \a_1$ and 
$\k_a(\om) - \a_1$ are respectively represented in blue and green. The unfolded curve $\k_a(\om)$ 
is not analytic at the boundaries of the Brillouin zones since it is represented in 
the case of real frequency $\om$. This unfolded curve $\k_a(\om)$ 
becomes analytic for frequency $\om$ with positive imaginary part.\label{figfold}
}
\end{figure}

For the given complex frequency $\om$ and the unit vector $\e_\k$ fixed in $S_\x^+$, 
the function $\dbtilde{G}(\k,\om)$ has a discrete set of poles with respect to the 
variable $\ks$, the number of poles could either infinite or zero in the case of a 
bandgap.
Let $\ks_p(\om, \e_\k)$ be the poles of the modulus $\ks$ of the wavevector 
in the direction $\e_\k$. 
These poles correspond to Bloch modes with complex wavevector $\ks_p(\om,\e_\k) \, \e_\k$ 
associated to the complex frequency $\om$ (see figure \ref{figDL}), and can have 
positive or negative imaginary part since it has been proved in 
\cite{Tip00} that for non real frequency $\om$ no real Bloch wavevector exists. 
Moreover, they form a discrete set which could either infinite or zero, since for 
real frequency $\om$, one can prove that the line corresponding to the direction 
$\e_\k$ will intersect the union of the Brillouin zones a periodic number of 
times if the slope of this line is rational (which is either zero in a stop band 
or infinite). If the slope is irrational these intersection points after 
$\K$-translations to the first Brillouin zone formed a set that is dense in 
the first Brillouin zone, therefore the number of intersections in that 
case is again infinite or zero in a bandgap. This still holds by perturbation for frequency $\om$ with 
a small imaginary part. 
Since $\e_\k \cdot \x$ is positive, the integral over $\ks$ can be calculated by 
closing the line of real numbers by a semi-circle (with infinite radius) in the upper 
half complex plane\footnote{Notice that the Fourier transform of the source 
(\ref{sourceJ}) is analytic of $\k$, since the support of the source is included 
in the ball of radius $c t_0$, and its exponential behavior is $\exp[\pm i k c t_0]$.}. 
In that case, the set of Bloch wavevectors with positive imaginary part 
$\k_p^+(\om,\e_\k) = \ks_p^+(\om,\e_\k) \, \e_\k$ are picked up by closing the 
loop in the complex plane:
\begin{equation}
G(\x,t) = \dfrac{2 i \pi}{(2 \pi)^4} 
\displaystyle\int_{R_\eta} \!\! d\om \, \exp[ - i \om t] \displaystyle\int_{S_\x^+} d\e_\k 
\, \displaystyle\sum_p \exp[ i \k_p^+(\om,\e_\k) \cdot \x \, ] \, 
\text{Res}[\k_p^+(\om,\e_\k), \om] \, ,
\label{Gxtpoles}
\end{equation}
where $\text{Res}[\k_p^+(\om,\e_\k), \om]$ is the residue of 
$\ks^2\dbtilde{G}(\ks \e_\k,\om)$ at the pole $\ks = \ks_p^+(\om)$. 
Then the integral over all the directions $\e_\k$ in the hemisphere $S_\x^+$ is 
performed. As a result, all the complex poles $\ks_p^+(\om,\e_\k)$, corresponding to the 
Bloch wavevectors $\k_p^+(\om,\e_\k)$, are picked up and the whole isofrequency 
dispersion law at the complex frequency $\om$ is obtained: this isofrequency 
dispersion law is periodic with respect to the wavevector $\k$. Let $\k_0^+(\om, \e_\k)$ 
be this isofrequency dispersion law restricted to the First Brillouin zone $\Bri$ (see figure 
\ref{figfold}).
First, it is assumed that this isofrequency dispersion law is well-defined for 
for all direction $\e_\k$, i.e. that the function $\k_0^+(\om,\e_\k)$ exists 
for all direction $\e_\k$ in the unit sphere $S$. This assumption is 
correct for frequencies $\om$ small enough at which, according to homogenization 
theory \cite{BLP78}, the photonic crystal behaves like a homogeneous medium. 
Also, without loss of generality, no more than a single mode is assumed in each 
direction in the first Brillouin zone: in practice, a finite sum over several modes 
in the first Brillouin could be considered. 

Since the dispersion law $\k(\om)$ is $\reclat$-periodic, the integral over $S_\x^+$ 
and the sum over $p$  in the expression (\ref{Gxtpoles}) can be re-arranged as a 
periodic sum over the reciprocal lattice $\reclat$ and the 
unit sphere $S$: 
\begin{equation}
\begin{array}{l}
\displaystyle\int_{S_\x^+} d\e_{\k} 
 \displaystyle\sum_p \, \exp[ i \ks_p^+(\om) \, \e_\k \cdot \x \, ] \, 
\text{Res}[\ks_p^+(\om) \, \e_\k, \om] \\[4mm]
\qquad \qquad \qquad = \displaystyle\sum_{\K \in \reclat} 
\displaystyle\int_{S} d\e_{\k} \, 
\exp\big[i \{ \k_0^+(\om,\e_\k) + \K \} \cdot \x \big] 
\text{Res}\big[\k_0^+(\om,\e_\k) + \K , \om \big] \, .
\end{array}
\label{disp-per}
\end{equation}
Let the function $R_\# \big[\x,k_0^+(\om,\e_\k),\om \big]$ be defined by
\begin{equation}
R_\#\big[\x, \k_0^+(\om,\e_\k),\om\big] = \dfrac{i}{(2\pi)^3}
\displaystyle\sum_{\K \in \reclat} 
\exp\big[i \{ \k_0^+(\om,\e_\k) + \K \} \cdot \x \big] 
\text{Res}\big[\k_0^+(\om,\e_\k) + \K , \om \big] \, .
\label{Rper}
\end{equation}
This function has some properties of a Floquet-Bloch component: it is $\reclat$-periodic 
with respect to $\k_0^+(\om,\e_\k)$ and it satisfies the Bloch boundary conditions 
with respect to $\x$. With this notation, the expression (\ref{Gxtpoles}) of $G(\x,t)$
becomes 
\begin{equation}
G(\x,t) = 
\displaystyle\int_{R_\eta} \!\! d\om \, \exp[ - i \om t] 
\displaystyle\int_{S} d\e_{\k} \, 
R_\#\big[\x, \k_0^+(\om,\e_\k),\om\big] \, .
\label{GxtDL}
\end{equation}
Now, from the analyticity property (\ref{equiv-om}), the function 
$R_\#\big[\x, \k_0^+(\om,\e_\k),\om\big]$ must be analytic with respect to $\om$ 
in the domain Im$(\om)>0$ as soon as the expression above is valid. 
This suggests that the ``dispersion law'' $\k_0^+(\om,\e_\k)$ could be also analytic 
under the same conditions if the function $R_\#\big[\x, \k_0^+(\om,\e_\k),\om\big]$ 
could be ``inverted''. In this aim, the Bloch boundary condition is used: for $\a$ 
in the lattice $\lat$ of the photonic crystal, the expression (\ref{Rper}) implies
\begin{equation}
R_\#\big[\x + \a, \k_0^+(\om,\e_\k),\om\big] = 
\exp\big[i \k_0^+(\om,\e_\k) \cdot \a \big] R_\#\big[\x, \k_0^+(\om,\e_\k),\om\big] \, .
\label{invert}
\end{equation}
Hence the function $\exp\big[i \k_0^+(\om,\e_\k) \cdot \a \big]$ is analytic 
as soon as $R_\#\big[\x, \k_0^+(\om,\e_\k),\om\big]$ does not vanish. Now, for 
$\k_0^+(\om,\e_\k)$ in the first Brillouin zone, the function 
$\exp\big[i \k_0^+(\om,\e_\k) \cdot \a \big]$ can be uniquely inverted and thus 
it is reasonable to consider that the ``dispersion law'' $\k_0^+(\om,\e_\k)$ 
is analytic with respect to $\om$. However, it is stressed that all these arguments 
are valid under the following conditions: the function $\k_0^+(\om,\e_\k)$ must exist 
for all direction $\e_\k$ in the unit sphere $S$ and must remain in the First 
Brillouin zone $\Bri$. These conditions are met for frequencies $\om$ small enough. 

The arguments for the analyticity of $\k_0^+(\om,\e_\k)$ presented above seem to fail 
when $\k_0^+(\om,\e_\k)$ reaches the contour of the first Brillouin zone. This is not 
surprising since, by construction, $\k_0^+(\om,\e_\k)$ corresponds to the isofrequency 
in the First Brillouin zone, and thus results from a folding of the dispersion law (see figure \ref{figfold}). 
However, the dispersion law $\k_0^+(\om,\e_\k)$ can be ``unfolded'' and the 
function (\ref{Rper}) can expressed as 
\begin{equation}
\begin{array}{ll}
R_\#\big[\x, \k_0^+(\om,\e_\k),\om\big] 
& = \dfrac{i}{(2\pi)^3}
\displaystyle\sum_{\K \in \reclat} 
\exp\big[i \{ \k_a^+(\om,\e_\k) + \K \} \cdot \x \big] 
\text{Res}\big[\k_a^+(\om,\e_\k) + \K , \om \big] \\[4mm]
& = R_\#\big[\x, \k_a^+(\om,\e_\k),\om\big] \, ,
\end{array}
\label{Ranalytic}
\end{equation}
where the function $\k_a^+(\om,\e_\k)$ is the \textbf{unfolded dispersion law} (see figure \ref{figfold}). 
This expression (\ref{Ranalytic}) is just a re-arrangement of the series in 
(\ref{Rper}) since the unfolded dispersion law $\k_a^+(\om,\e_\k)$ is defined 
for high frequencies as a translation of dispersion law in the first Brillouin zone 
by a vector in the reciprocal lattice $\reclat$: for all frequency $\om$, there 
exists a vector $\K$ in $\reclat$ such that 
\begin{equation}
\k_a^+(\om,\e_\k) = \k_0^+(\om,\e_\k) + \K \, .
\label{unfold}
\end{equation}
This unfolded dispersion law can uniquely defined by analytic continuation at the 
boundaries of the Brillouin zones. 

Now, the solution of Maxwell's equations can be expressed using the 
unfolded dispersion law:
\begin{equation}
G(\x,t) = 
\displaystyle\int_{R_\eta} \!\! d\om \, \exp[ - i \om t] 
\displaystyle\int_{S} d\e_{\k} \, 
R_\#\big[\x, \k_a^+(\om,\e_\k),\om\big] \, .
\label{GxtDLA}
\end{equation}
Again, the function $R_\#\big[\x, \k_a^+(\om,\e_\k),\om\big]$ must be analytic 
in the domain of frequencies Im$(\om)>0$. Then, the unfolded dispersion law 
$\k_a^+(\om,\e_\k)$ can be extracted from $R_\#\big[\x, \k_a^+(\om,\e_\k),\om\big]$ 
using the argument (\ref{invert}): hence it obtained that 
\begin{equation}
\exp\big[i \{ \k_a^+(\om,\e_\k) + \K \} \cdot \a \big] = 
\exp\big[i \{ \k_0^+(\om,\e_\k) + \K \} \cdot \a \big] \, ,
\end{equation}
which is consistent with (\ref{unfold}), but now the inversion of the 
exponential function must be done in the way that preserves $\k_a^+(\om,\e_\k)$ 
analytic when it spans the whole reciprocal space. Thus the unfolded dispersion law 
$\k_a^+(\om,\e_\k)$ appears as the analytic 
continuation from the small frequencies $\om$ of $\k_0^+(\om,\e_\k)$ in 
the first Brillouin zone. \\

\noindent
\textbf{Discussion.} Arguments based on the causality principle have been 
proposed to support that the unfolded dispersion law 
$\k^+(\om) \equiv \k_a^+(\om,\e_\k)$ is an 
analytic function of the frequency in the domain of complex frequencies $\om$ 
with positive imaginary part. This dispersion law $\k^+(\om)$ has been defined 
with a positive imaginary part. A similar dispersion law $\k^-(\om)$ with 
a negative imaginary part could be defined using, instead of the hemisphere 
$S_\x^+$ defined by (\ref{Splus}), the hemisphere
\begin{equation}
S_\x^- = \big\{ \e_\k \in S \, | \, \e_\k \cdot \x < 0 \, \big\} \, . 
\label{Smoins}
\end{equation}
Indeed, in that case, the step from equation (\ref{Gxtspherique}) to equation 
(\ref{Gxtpoles}) is performed by closing 
the real axis by a semi-circle in the lower half complex plane of number $\ks$, 
leading to pick up the poles with negative imaginary parts. 
As a consequence, it is found that in the domain of frequencies $\om$ with 
positive imaginary parts two distinct analytic dispersion laws $\k^\pm (\om)$ exist, 
with $\k^+(\om) = - \k^-(\om)$. 

For small frequencies, the dispersion law $\k^\pm(\om)$ is well-defined 
for all direction $\e_\k$ in the unit sphere $S$. By analytic continuation, 
the dispersion law $\k^\pm(\om)$ appears to be well-defined for all frequencies $\om$
and for all directions $\e_\k$, which could be considered as surprising. Indeed, for real 
frequencies and real wavevectors the periodicity of photonic crystal implies the 
presence of bandgaps and, more frequently, of stop bands (i.e. the absence of 
Bloch modes for certain directions $\e_\k$). However, when considered in the complex plane, 
it appears that one can find a complex wavevector $\k$ for all frequency $\om$. 

The present conclusions have been rigorously proved and numerically checked in the 
one-dimensional case in the reference \cite{Liu2013}. In particular, it has been 
shown that the wavenumber $\ks(\om)$ is an analytic function with respect to 
the frequency $\om$ in the domain Im$(\om)>0$ and that its imaginary part cannot 
vanish (passivity requirement). Here, a similar result has been found since the 
two analytic unfolded dispersion laws $\k^\pm(\om)$ have keep the same sign for 
their imaginary part: hence the wavevectors $\k^\pm(\om)$ cannot vanish. 
In the one-dimensional case \cite{Liu2013}, all these results have been confirmed 
numerically, for instance by checking the validity of the Kramers-Kronig relations. 

It is stressed that the arguments presented in this section remain valid in 
the case of dispersive and absorptive photonic crystals since it preserves 
the analytic nature with respect to the frequency. 

Finally, the reciprocal dispersion law $\om(\k)$ has not been considered 
in this last section. Indeed, complex Bloch wavevector cannot be directly 
introduced with Fourier transform since they imply exponential growing 
in the integrals. However, from the analytic properties (\ref{equivalence}) of 
$\dbtilde{G}(\k,\om)$, it can be expected that a well-defined dispersion law 
$\om(\k)$ could have analytic properties as soon as Im$(\k) \neq 0$.

\section{Conclusion\label{sec8}}

This chapter has been focused on fundamental definitions and properties of 
dispersion law and group velocity in photonic crystals, including 
illustrations with numerical examples. This review has shown that numerous 
questions need to be investigated in the future. The numerical computation 
of the dispersion law becomes very challenging when the dispersion and 
absorption are introduced. The techniques based on the introduction of the 
auxiliary fields \cite{Tip98,Tip00,GT10} have been developed and numerically 
implemented \cite{Fan10,Bru16} and are now the basic tool for the emerging topic
of quasi-normal modes in photonics \cite{lalanne2018,Wei2018}. It is stressed 
that these numerical tools use only partial extension of Maxwell's equations 
where the solely dispersion is removed. A challenging question will be to 
implement the full extension of Maxwell's equations \cite{Tip98,Tip00,GT10} 
which is associated to a selfadjoint time-independent operator. In particular 
such an extension may bring an answer to the open questions of the completeness and 
the normalization of the quasi-normal mode expansions, as well as the link between 
the complex resonances of the quasi-normal mode expansions and the real spectrum 
of the augmented selfadjoint operator. Another open question is the analytic 
structure of the dispersion law. Simple arguments based on the causality 
principle have been proposed to support some analyticity properties, but 
rigorous investigations remain definitely necessary. Also, the numerical 
calculations of the dispersion law in dispersive and absorptive photonic 
crystals \cite{Bru16} have shown that the resonances associated with the 
first Brillouin zone contour $\Gamma X M \Gamma$ do not form the contour of 
the Bloch spectrum: the presence of resonances outside this closed path formed 
by the resonances of the contour has been highlighted. 
These preliminary investigations show the potential richness of the vast field of 
spectral theory of non-selfadjoint operators. Again, the full extension of 
dispersive and absorptive Maxwell's equations \cite{Tip98,Tip00,GT10} might 
be a starting point to explore the spectral theory of non-selfadjoint 
operators.

The modeling of photonic crystals as effective homogeneous media received 
a keen interest of the community, which led to important contributions in the 
homogenization theory. Homogenization is an old subject, which dates back to the 
work by Lord Rayleigh on quasi-static analysis of periodic non dissipative 
structures\cite{rayleigh1892}. Physicists and mathematicians have used various 
approaches to replace a periodic structure by an effective medium in the 
long-wavelength limit with semi-analytical multipole Rayleigh expansions 
in the dipole approximations \cite{McPhedran1996}, plane wave expansions 
\cite{Halevi1999}, or asymptotic multiple scale expansions techniques 
\cite{BLP78,Guenneau2000}, and a variety of variational techniques such as 
the compensated compactness of Tartar and we refer the reader to the book by 
Milton for a review of low frequency homogenization theories in the 
composite community \cite{Milton02}. Interestingly, if one adds further 
corrections to the usual averaged properties of photonic crystals, which is 
the consecrated high-order homogenization \cite{Liu2013}, it is necessary to 
add effective tensors of magneto-optic coupling and permeability to the usual 
tensor of effective permittivity in order to accurately describe the effective medium.
Another pitfall of classical homogenization in photonic crystals is the effect 
of the boundaries on effective properties 
\cite{Silveirinha2006,Pierre2008,Smigaj2008,Markel2013}. 
These works touch upon concept of non-locality in homogenization of finite 
photonic crystals with moderate \cite{Pierre2008,Smigaj2008,Markel2013} and high 
\cite{Silveirinha2006} contrast. Frequency dispersion in effective properties of 
high contrast photonic crystals has been also investigated in 
\cite{Felbacq1997,Silveirinha2006,Silveirinha2007}. In order to extract the unusual 
effective parameters of photonic crystals and metamaterials at any frequency, one 
can also use some numerical approaches such as the retrieval method, which 
amounts to fitting the reflection and transmission coefficients of a given complex 
medium with those of an effective medium through a numerical optimization 
procedure \cite{Koschny2003,Menzel2008}. Another popular method to compute the 
effective properties of a periodic structure is a homogenization technique in which 
macroscopic fields are determined via averaging of the local fields obtained from a 
full-wave electromagnetic simulation \cite{Smith2006}. In the same vein, the high-frequency 
homogenization allows to reconstruct dispersion curves and associated Bloch waves 
through a procedure based on numerical field averaging in a periodic cell at any 
frequency \cite{Craster2010}. 

Finally, the most recent investigations on the bands in photonic crystals 
focus on topological insulators. For certain ranges of frequency, these structures 
behave as insulators in their bulk but allow edge states to propagate along a line 
defect of the photonic crystal. These edge states as surface waves are transversely 
localized to the defect.
Compared to other insulator structures, the main feature of topological 
insulators \cite{Chan-14,Kan-05,Rag-08,Recht-13} is that the edge states 
have the particularity to be \emph{topologically protected} or, in other words, 
very robust to perturbations of the line defect that do not break the bandgap 
(as for instance local perturbation of the interface of the defect). Moreover these 
edge states do not backscatter under such perturbations. 
Examples of topological insulators are given by graphene \cite{Weinstein-2018} 
or topological graphene \cite{Recht-13,Wein-2017}. 
Indeed, based on the symmetry of the two dimensional honeycomb structure 
of the graphene crystal (which is invariant by rotation of angle $2\pi/3$ and 
inversion), one can show 
that certain couple of dispersion curves (see for graphene \cite{Wall-47,Weinstein-2018} 
and for photonic graphene \cite{Wein-2017}) degenerate at the vertices of the 
first Brillouin zone (which is here hexagonal)
where they cross conically on points referred in the literature as Dirac points. 
Perturbing the dispersion curves at a Dirac point 
with a line defect that breaks the $\mathcal{P}\mathcal{T}$ symmetry (i.e. the 
composition of parity-inversion and time-reversal symmetries) of the crystal allows 
to open a gap (that could be only a local gap for the case of photonic 
graphene see \cite{Wein-2017}). In addition, such a defect ensures the 
existence of topology protected edge states which are localized in this gap 
\cite{Wein-2017,Weinstein-2018}. 


\end{document}